\documentclass[submission, Phys]{SciPost}
\usepackage{verbatim}
\usepackage{amsmath}
\usepackage{capt-of}
\usepackage{amssymb}
\usepackage{graphicx}
\usepackage{xcolor}
\usepackage{times}
\usepackage{soul}

\usepackage{algorithm}
\usepackage{multicol}
\usepackage{algpseudocode}

\def \beq{\begin{equation}}
\def \eeq{\end{equation}}
\def \barray{\begin{eqnarray}}
\def \earray{\end{eqnarray}}

\font\numbers=cmss12
\font\upright=cmu10 scaled\magstep1
\def\stroke{\vrule height8pt width0.4pt depth-0.1pt}
\def\topfleck{\vrule height8pt width0.5pt depth-5.9pt}
\def\botfleck{\vrule height2pt width0.5pt depth0.1pt}
\def\Zmath{\vcenter{\hbox{\numbers\rlap{\rlap{Z}\kern
0.8pt\topfleck}\kern 2.2pt
                   \rlap Z\kern 6pt\botfleck\kern 1pt}}}
\def\Qmath{\vcenter{\hbox{\upright\rlap{\rlap{Q}\kern
                   3.8pt\stroke}\phantom{Q}}}}
\def\Nmath{\vcenter{\hbox{\upright\rlap{I}\kern 1.7pt N}}}
\def\Cmath{\vcenter{\hbox{\upright\rlap{\rlap{C}\kern
                   3.8pt\stroke}\phantom{C}}}}
\def\Rmath{\vcenter{\hbox{\upright\rlap{I}\kern 1.7pt R}}}

\begin{document}

\begin{center}{\Large \textbf{
Molecular dynamics simulation of entanglement spreading in generalized hydrodynamics
}}\end{center}

\begin{center}
M\'arton Mesty\'an\textsuperscript{1*},
Vincenzo Alba\textsuperscript{2}
\end{center}

\begin{center}
{\bf 1} International School for Advanced Studies (SISSA) and INFN,\\
Via Bonomea 265, 34136, Trieste, Italy \\
{\bf 2} Delta Institute for Theoretical Physics, University of Amsterdam,\\
Science Park 904, 1098 XH Amsterdam, the Netherlands
\\
* mestyan@fmf.uni-lj.si
\end{center}

\begin{center}
\today
\end{center}

\section*{Abstract} 

{\bf

We consider a molecular dynamics method, the so-called flea gas for computing the evolution of entanglement after inhomogeneous quantum quenches in an integrable quantum system.
In such systems the evolution of local observables is described at large space-time scales by the Generalized Hydrodynamics approach, which is based on the presence of stable, ballistically propagating quasiparticles.
Recently it was shown that the GHD approach can be joined with the quasiparticle picture of entanglement evolution, providing results for entanglement growth after inhomogeneous quenches.
Here we apply the flea gas simulation of GHD to obtain numerical results for entanglement growth. We implement the flea gas dynamics for the 
gapped anisotropic Heisenberg XXZ spin chain, considering quenches from globally homogeneous and 
piecewise homogeneous initial states. 
While the flea gas method applied to the XXZ chain is not exact even in the scaling limit (in contrast to the Lieb--Liniger model), it yields a very good approximation of analytical results for entanglement growth in the cases considered.
Furthermore, we obtain the {\it full-time} dynamics of the
mutual information after quenches from inhomogeneous settings, for which 
no analytical results are available. 
}


\section{Introduction}
\label{intro}

In the last decade, the study of isolated quantum many-body 
systems out of equilibrium provided new insights on the deep interplay between 
entanglement and thermodynamics, shedding 
new light on the fundamental question how statistical ensembles 
emerge from the out-of-equilibrium dynamics after a {\it quantum 
quench}~\cite{ge-15,R08,ef-16,ViRi16,CQA16,DKPR16}. 

\begin{figure}[t]
\begin{center}
\includegraphics[width=0.75\linewidth]{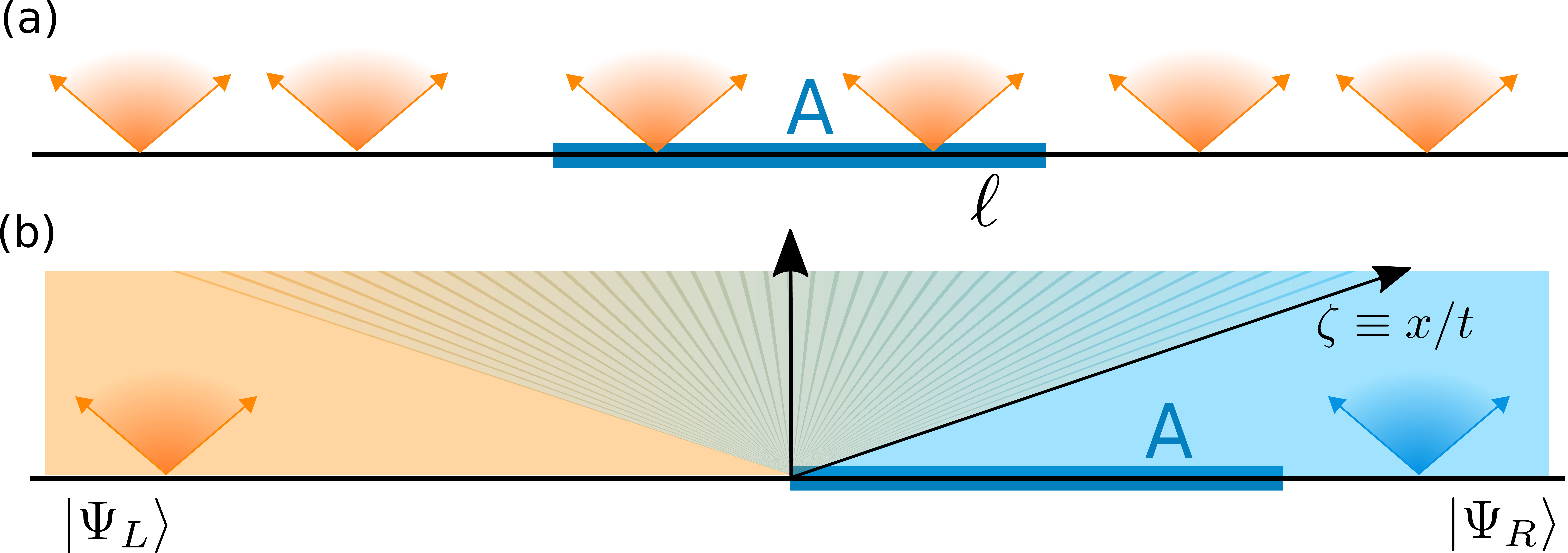}
\end{center}
\caption{ Dynamics of the entanglement entropy after a quench from a 
 homogeneous (in (a)) and a piecewise homogeneous initial condition (in (b)). 
 In both cases the entanglement dynamics is due to the 
 ballistic propagation of pairs of entangled particles (shaded cones). 
 In (a) the quasiparticles entanglement entropy is the thermodynamic 
 entropy of the GGE that describes the steady state. 
 In (b) the inhomogeneous 
 initial state is obtained by joining two homogeneous systems (left and right) 
 in the states $|\Psi_L\rangle$ and $|\Psi_R\rangle$. Entangled pairs are produced 
 in the bulk of the two chains. A lightcone spreads 
 from the interface between them. For each value of $\zeta\equiv x/t$ 
 the system relaxes locally to a GGE. 
 The entanglement entropy is obtained propagating the entropy of the GGEs 
 describing the bulk of the left and right chains, i.e., for $\zeta\to\pm\infty$. 
}
\label{fig0}
\end{figure}

Integrable models offer an ideal setting  for understanding generic features of 
the entanglement dynamics after a quantum quench~\cite{FaCa08,CCsemiclassics,dmcf-06,lauchli-2008,ep-08,hk-13,Gura13,Fago13,ctc-14,nr-14,buyskikh-2016,cotler-2016,kctc-17,MBPC17,mkz-17,p-18,fnr-17,ckt-18,d-17,BTC:beyondpairs,BC:beyondpairs2,LS:CFT,CLM:minimalcut,ABGH:CFT,LM:CFT,AlCa17,nahum-17,BKP:entropy,MM}. 
Indeed, for integrable systems the dynamics of 
entanglement-related quantities can be 
understood within the so-called quasiparticle picture~\cite{CCsemiclassics}. Here we focus on the 
out-of-equilibrium dynamics of the entanglement entropy (von Neumann 
entropy), which is defined as~\cite{area,amico-2008,calabrese-2009,laflorencie-2016} 
\begin{equation}
S=-\textrm{Tr}\rho_A\ln\rho_A, 
\end{equation}
with $\rho_A$ the reduced density matrix of a macroscopic subsystem $A$ 
(see Fig.~\ref{fig0} for a one-dimensional setup). 
In the quasiparticle picture, pairs of entangled quasiparticles are produced after 
the quench. As these pairs propagate  ballistically, they entangle 
larger regions of the system (see Fig.~\ref{fig0} (a)). At a given time after 
the quench, the von Neumann entanglement entropy is the sum of the individual contributions coming from each pair 
that is shared between $A$ and its complement.  This picture 
has been explicitly verified in free-fermion models~\cite{FaCa08}. 
It has been shown recently that it holds true also in the presence of 
interactions~\cite{AlCa17}. 

The quasiparticle prediction for the entanglement entropy of subsystem $A$ of length 
$\ell$ after a quench in generic integrable systems reads as~\cite{AlCa17} 
\begin{equation}
\label{semi-i}
S(t)= \sum_\alpha\Big[ 2t\!\!\!\!\!\!\int\limits_{\!2|v_{\alpha,\lambda}|t<\ell}\!\!\!\!\!\!
  d\lambda   |v_{\alpha,\lambda}| s_{\alpha,\lambda}+\ell\!\!\!\!\!\!\int\limits_{2|v_{\alpha,\lambda}|t>\ell}\!\!\!\!\!\!
d\lambda  s_{\alpha,\lambda}\Big]. 
\end{equation}
Here the index $\alpha$ labels the different types of quasiparticles 
present in the model, and $\lambda$ is the so-called rapidity, which 
distinguishes different modes of the same type of quasiparticles. 
The quasiparticle picture is built on two important 
ideas. First, at long times the entanglement entropy between $A$ and its complement 
coincides with the thermodynamic entropy of the statistical 
ensemble that describes local steady-state properties in $A$. For 
generic translationally invariant steady states, this ensemble is a Generalized Gibbs 
Ensemble~\cite{R08,ef-16,ViRi16,CQA16,DKPR16,rigol-2007,wouters-2014,pozsgay-2014, IDWC15} 
(GGE). Second, the velocity of the entangling quasiparticles are the group 
velocities of the particle-hole excitations~\cite{BEL:lightcone} over the GGE steady state. A crucial observation is that, in contrast with free-fermion models, 
$v_{\alpha,\lambda}$ depends on the GGE describing the steady state 
after the quench (or equivalently on the pre-quench initial state), 
and it is ``dressed'' by the interactions. For generic lattice 
models with local interaction the velocities are finite, i.e, for any $\alpha,\lambda$, $|v_{\alpha,\lambda}|<\infty$.
The function $s_{\alpha,\lambda}$, i.e., the entropy carried by quasiparticle 
pairs of type $\alpha$ in mode $\lambda$, is the contribution 
of these quasiparticles to the thermodynamic entropy of the GGE describing 
subsystem $A$.

The approach of Ref.~\cite{AlCa17} has been generalized in~\cite{ac-17b,alba-2018,mestyan-2018} 
to calculate the R\'enyi entropies in the steady-state, whereas calculating their 
full-time dynamics is a challenging open problem. Remarkably, the quasiparticle picture 
allows to obtain the dynamics of the logarithmic negativity~\cite{ctc-14,alba-2018a}, which is a 
proper entanglement measure for mixed states.  
%
%

Recently, there has been a growing interest in understanding the entanglement 
dynamics after quenches from piecewise homogeneous initial  states. 
In the standard setup (see Fig.~\ref{fig0} (b)) two homogeneous chains 
in a different state  ($L,R$ in Fig.~\ref{fig0}) are 
joined together at $t=0$. One then studies the ensuing dynamics under an 
integrable globally homogeneous Hamiltonian. During the time evolution a 
lightcone spreads from the interface between the chains. 
For typical initial states, in the limit of long times and 
large distances $x$ from the origin, the system  reaches at each fixed ray $\zeta\equiv x/t$  (see  Fig.~\ref{fig0}), a Local 
Quasi Stationary State~\cite{BF16} (LQSS) 
which is described by a GGE. These $\zeta$-dependent GGEs can be described 
analytically within  the Generalized Hydrodynamics (GHD) formalism~\cite{BCDF16,CaDY16}.  
Furthermore, by combining the quasiparticle picture with 
the GHD approach~\cite{Alba17,alba-2018,BFPC,alba-2019} 
it is possible, in principle, to generalize the quasiparticle picture~\cite{AlCa17} to 
inhomogeneous settings. However, actually calculating the full-time entanglement 
dynamics in this case is a demanding task. The main difficulty is that, unlike in homogeneous 
quenches, the trajectories of the quasiparticles 
are not straight lines. Explicit results are available only in the 
short time limit~\cite{alba-2019} $t/\ell\to 0$ and the long time limit $t/\ell\to\infty$.

In order to overcome the above difficulties, we use a mapping between the GHD and the ``molecular dynamics'' 
  of a system of classical particles called flea gas~\cite{DDKY17}. 
 This mapping allows one to obtain the out-of-equilibrium dynamics in 
quantum integrable systems by performing classical 
simulations~\cite{jerome-2018,bdwy-18,DoYo17a}. 
So far, the flea gas method has been employed only to integrable field theories, such as the Lieb-Liniger gas, but not to lattice models. In this paper, we discuss a generalization of the 
flea gas approach to  the spin-$1/2$ anisotropic XXZ chain. An important remark is that, unlike 
for the Lieb-Liniger model~\cite{DDKY17}, for the XXZ chain it is not 
straightforward to show analytically that the flea gas dynamics is 
fully equivalent to the GHD.  
In general for the XXZ chain the flea gas dynamics is expected to be 
different from the GHD. Our results suggest that that deviations are 
present, although much larger systems would be needed to ensure 
that the system is in the scaling limit. Still, we observe that deviations 
of the flea gas from the GHD are small. This results in a surprisingly good 
agreement between the flea gas and the GHD for the dynamics of local 
observables and the entanglement entropy. 

Our main goal is to show that with moderate computational effort the flea gas framework 
allows  one to compute accurate numerical predictions for the {\it full-time} entanglement 
dynamics after quenches from arbitrary inhomogeneous initial conditions, 
provided that the Generalized Hydrodynamics approach can be applied. 

The flea gas algorithm requires to know the {\it bare} quasiparticle velocity 
and the scattering matrix between quasiparticles, which are easily obtained for a generic integrable model.
The third and only non-trivial input to this method is the initial condition of 
the flea gas dynamics. For quenches from homogeneous initial states, 
the initial condition of the flea gas dynamics is given by the GGE describing the long time limit after the quench. The idea is that at comparatively 
short times after the quench the quasiparticles are described by the GGE density. 
For piecewise homogeneous setups (see, e.g., Fig.~\ref{fig0} (b)), 
the initial condition is given by the GGEs describing each homogeneous chain 
at $t=0$ in the hydrodynamic timescale.  

We stress that it is also necessary to know the structure of quantum 
correlations in the initial states, i.e., how entanglement 
is shared among the quasiparticles. Here we restrict 
ourselves to the situation in which only entangled {\it pairs} are present. 
Other situations, for instance the case of entangled 
``triplets'', have been considered, at least in free 
models~\cite{BTC:beyondpairs,BC:beyondpairs2}.

As a benchmark of the method, we show that for quenches from homogeneous states 
our numerical results are in perfect agreement with~\eqref{semi-i}. Moreover, 
for quenches from inhomogeneous initial states, in the limit $t,\ell\to\infty$ 
with $t/\ell\ll1$ our results confirm the analytical predictions in 
Ref.~\cite{alba-2019}. Finally, to show the 
versatility of the method we provide results for the full-time dynamics 
of the entanglement entropy and of the mutual information after a quench 
from inhomogeneous settings, which are not easily accessible 
analytically~\cite{alba-2019}. 

The manuscript is organized as follows. In Section~\ref{sec-quenches} we introduce the 
XXZ chain and the quenches considered in this study. In Section~\ref{tba-xxz} we discuss 
the Bethe ansatz treatment of generic thermodynamic ensembles. 
 The thermodynamic Bethe ansatz framework (TBA)
is introduced in Section~\ref{sec-tba},  which is followed by the description
of steady states after homogeneous quenches in Section~\ref{sec-tba-quench}
and a summary of the GHD approach for inhomogeneous quenches in Section \ref{ghd}. 
In Section~\ref{sec-flea} we introduce the flea gas method, 
its implementation for the XXZ chain (see Section~\ref{sec-flea-xxz}), and the 
calculation of entanglement-related quantities (see Section~\ref{sec-flea-ent}). Our 
numerical results are discussed in Section~\ref{sec-ent-num}. In Section~\ref{sec-ent-h} we  benchmark the method for homogeneous quenches. In 
Section~\ref{sec-ent-in} we provide results for 
entanglement entropy after quenches from piecewise homogeneous initial states. Finally, 
in Section~\ref{sec-ent-mi} we discuss the mutual information. 
Section~\ref{sec-con} concludes the article by mentioning some interesting  future directions.

\section{Model and quenches}
\label{sec-quenches}

The  flea gas method that we propose for calculating the entanglement dynamics 
is expected to work for generic interacting integrable models, both on the lattice and 
in the continuum. However, here we provide results only for a prototypical 
lattice model, the spin-$1/2$ XXZ chain, which describes a system of 
interacting spins on a ring, and it is defined by the Hamiltonian
\begin{equation}
\label{xxz-ham}
H=\sum_{i=1}^L\frac{1}{2}(S_i^+S_{i+1}^-+S_{i}^-S_{i+1}^+)+
\Delta\sum_{i=1}^L S_i^zS_{i+1}^z. 
\end{equation}
Here $S_i^{+,-,z}$ are spin-$1/2$ operators, and $\Delta$ is the 
so-called anisotropy parameter.  We restrict ourselves 
to the region with $\Delta>1$, where the system is gapped in 
the thermodynamic limit, although the method can be applied for 
$\Delta\le1$ as well. We impose periodic boundary conditions 
by identifying sites  $1$ and $L+1$. 
 We construct our initial states by joining two homogeneous 
 blocks that are prepared in either the translationally invariant 
 tilted N\'eel state or in the translationally invariant  Majumdar-Ghosh 
(dimer) state. The method is applicable, 
in principle, to any low-entangled initial state. 

The translationally invariant tilted N\'eel state is denoted as $|\mathrm{N},\theta\rangle$, 
and it is obtained by rotating the N\'eel state $\left|
\uparrow\downarrow\uparrow\dots\right\rangle$ around the $\hat z$ 
axis and making it translationally invariant, i.e., 
\begin{equation}
  \begin{split}
  \quad|\mathrm{N},\theta\rangle =  \left( \frac{1+\mathcal T}{\sqrt 2} \right)\ &\,\bigg\{ [\cos(\theta/2)\left|\uparrow\right\rangle +  
 i\sin(\theta/2)\left|\downarrow\right\rangle]\, \otimes  \\
 &\, \otimes [\sin(\theta/2)
\left|\uparrow\right\rangle -i \cos(\theta/2)\left|\downarrow\right\rangle] 
\bigg\}^{\otimes L/2}.
\end{split}
\label{eq:tiltedNeel}
\end{equation}
Here $\theta$ is the tilting angle and $\mathcal T$ is the one site translation 
operator to the right. The  N\'eel state is 
recovered for $\theta=0$. Similarly, the translationally invariant dimer 
state $|\mathrm{D}\rangle$ is defined as 
\begin{equation}
|\mathrm{D}\rangle = \left( \frac{1+\mathcal T}{\sqrt 2} \right)\left(\frac{\left|\uparrow\downarrow
\right\rangle - \left|\downarrow\uparrow\right\rangle }
{\sqrt 2} \right)^{\otimes L/2}.
\label{eq:dimer}
\end{equation}
In the homogeneous setup (Fig.~\ref{fig0} (a)), the chain is prepared in one of 
the states ~\eqref{eq:tiltedNeel} or~\eqref{eq:dimer} at $t=0$, and the system 
is let to evolve under~\eqref{xxz-ham}. In the inhomogeneous case (Fig.~\ref{fig0} (b)) 
we consider quenches from the initial state  $|\Psi_0\rangle=|\mathrm{N},\theta\rangle\otimes 
|\mathrm{D}\rangle$. 

\section{Bethe ansatz description of thermodynamic macrostates in the\\ XXZ~chain}
\label{tba-xxz}

Here we introduce the thermodynamic Bethe ansatz 
(TBA) treatment of the XXZ chain~\cite{Taka99}, focusing on the 
 features that are needed in the implementation of the flea gas method. First,  we 
summarize the general TBA framework in Section~\ref{sec-tba}.  Then we report the TBA 
  description of the steady states after the considered homogeneous quenches in Section~\ref{sec-tba-quench}. 
Finally, we summarize the generalized hydrodynamics (GHD) framework for quenches from 
inhomogeneous states in Section~\ref{ghd}.  

\subsection{Thermodynamic Bethe Ansatz (TBA)}
\label{sec-tba}

The XXZ chain is solved by the Bethe ansatz~\cite{Taka99}, which allows one to 
construct the eigenstates of~\eqref{xxz-ham}. In the Bethe ansatz, the eigenstates
  are constructed with respect to the reference state with all spins up $|\uparrow\uparrow\cdots\uparrow\rangle$.
  Since the total magnetization
  $\sum_{j} S_{j}^{z}$ commutes with \eqref{xxz-ham}, the eigenstates 
  are characterized by the total number $N$ of down spins, which is 
  a good quantum number  of the state. We refer to $N$ as the number of particles. 
  In this study we focus 
on the thermodynamic limit, i.e., the limit $L,N\to\infty$, with particle 
density $N/L$ fixed. 

A distinctive feature of integrable models is  that their eigenstates
can be interpreted as a collection of 
well-defined, i.e., having infinite lifetime, quasiparticles. For generic 
integrable models the quasiparticles are labelled by a set of real parameters 
 $\{\lambda_{\alpha,j}\}_{\alpha,j}$, which are called rapidities. 
For the XXZ chain at $\Delta>1$ one has $\lambda_{\alpha,j}\in[-\pi,\pi]$. 
In general, there can be different species of 
quasiparticles. These are distinguished by the integer index 
$\alpha$. The total number of species depends on $\Delta$. For instance, 
at $\Delta\ge 1$ there is an infinite number of them. 
Quasiparticles with $\alpha=1$ correspond to magnon-like excitations, whereas 
for $\alpha>1$ they are bound states of $\alpha$ magnons ($\alpha$-strings~\cite{Taka99}). 

In the thermodynamic limit it is impossible to consider the individual rapidities 
$\lambda_{\alpha,j}$ of the quasiparticles. Instead, the standard TBA 
framework~\cite{Taka99} uses the density of quasiparticles in rapidity space 
$\rho_{\alpha,\lambda}$, which are real functions of $\lambda$ for each species $\alpha$. 
One can also define the hole density $\rho_{\alpha,\lambda}^{\scriptscriptstyle
(\mathrm{h})}$  as the density of unoccupied states in rapidity space. 
Another important quantity is the total density of states 
$\rho_{\alpha,\lambda}^{\scriptscriptstyle(\mathrm{t})}=
\rho_{\alpha,\lambda}+\rho_{\alpha,\lambda}^{\scriptscriptstyle(\mathrm{h})}$.
For a generic integrable model, this is a non-trivial function of $\lambda$, 
whereas for non-interacting systems the total density is a constant, reflecting that 
the rapidities are equally spaced. For the following, it is also convenient to define the 
filling functions $\vartheta_{\alpha,\lambda}$ and the functions 
$\eta_{\alpha,\lambda}$ as 
\begin{equation}
\label{fill-eta}
\vartheta_{\alpha,\lambda}\equiv\frac{\rho_{\alpha,\lambda}}
{\rho_{\alpha,\lambda}^{(\mathrm{t})}},
\quad\eta_{\alpha,\lambda}\equiv\frac{\rho_{\alpha,
\lambda}^{(\mathrm{h})}}{\rho_{\alpha,\lambda}} .
\end{equation} 
The densities $\rho_{\alpha,\lambda}$ and $\rho_{\alpha,\lambda}^{\mathrm{h}}$ are constrained by the Bethe equations arising from the periodic boundary conditions.
 In the thermodynamic limit, the Bethe equations become the Bethe--Gaudin--Takahashi 
(BGT) equations \cite{Taka99}
\begin{equation}
\rho_{\alpha,\lambda} + \rho_{\alpha,\lambda}^{(\mathrm{h})} = 
a_{\alpha,\lambda} - \sum_{\beta=1}^{\infty} \int_{-\pi/2}^{\pi/2}\! 
d\mu\, T_{\alpha\beta}(\lambda-\mu) \rho_{\beta,\mu},
 \label{eq:BGT}
\end{equation}
where the functions $a_{\alpha,\lambda}$ are
\begin{equation}
a_{\alpha,\lambda} = \frac{1}{\pi} \frac{\sinh(\alpha 
\eta)}{\cosh(\alpha \eta)-\cos(2 \lambda)},
\label{eq:an}
\end{equation}
and $\eta\equiv\mathrm{arccosh}(\Delta)$. In~\eqref{eq:BGT}, the 
scattering matrix $T_{\alpha,\beta}$ is defined as 
\begin{equation}
  T_{\alpha,\beta}(\lambda) = (1-\delta_{\alpha\beta})a_{|\alpha-\beta|,\lambda} + 
  2a_{|\alpha-\beta|+2,\lambda} + \cdots + 2a_{\alpha+\beta-2,\lambda} 
  + a_{\alpha+\beta,\lambda},
  \label{eq:anm}
\end{equation}
where $a_{\alpha,\lambda}$ is the same as in~\eqref{eq:an}. The matrix $T_{\alpha,\beta}
(\lambda-\mu)$ encodes all the information about the scattering between 
quasiparticles of type $(\alpha,\lambda)$ and $(\beta,\mu)$, and it will be crucial 
in the implementation of the flea gas algorithm (see section~\ref{sec-flea}). 

In the TBA framework, any set of particle and hole densities 
$\rho_{\alpha,\lambda},\rho_{\alpha,\lambda}^{\scriptscriptstyle 
(\mathrm{h})}$ identifies a thermodynamic macrostate.  
All the information about thermodynamic expectation 
values of local and quasi local operators is encoded in the functions $\rho_{\alpha,\lambda}$. 
 These expectation values are determined by summing over the quasiparticles species 
and integrating over their rapidity. For instance, for the XXZ chain 
the energy of a macrostate identified by a set of densities $\rho_{\alpha,\lambda}$ 
reads~\cite{Taka99} 
\begin{equation}
\label{xxz-en}
\frac{E}{L}=\sum_\alpha\int_{-\pi/2}^{\pi/2} d\lambda \epsilon_{\alpha,
\lambda}\rho_{\alpha,\lambda}, \qquad\mathrm{with}\, \epsilon_{\alpha,
\lambda}= -\frac{\sinh{\eta}\sinh(\alpha\eta)}{\cosh(\alpha \eta)-\cos(2\lambda)}. 
\end{equation}
Besides the energy, integrable models have an infinite set of quasilocal charges that commute with the Hamiltonian. In the case of the XXZ model, these charges are obtained as the derivatives of the transfer matrix. In the thermodynamic limit, a 
conserved quasilocal charge ${\hat Q}$ is expressed as 
\begin{equation}
\frac{\langle \hat Q\rangle}{L}=\sum_\alpha\int^{\pi/2}_{-\pi/2}d\lambda{ 
q}_{\alpha,\lambda}\rho_{\alpha,\lambda}, 
\end{equation}
where ${q}_{\alpha,\lambda}$ is a known function, the density of the charge.

An important quantity that we will use is the bare 
velocity of the quasiparticles $v_{\alpha,\lambda}^{\mathrm{bare}}$. 
This is  the group velocity defined from the bare quasiparticles 
dispersion as 
\begin{equation}
\label{v-bare}
v_{\alpha,\lambda}^{\mathrm{bare}}\equiv
\frac{\epsilon'_{\alpha,\lambda}}{p'_{\alpha,\lambda}}
\qquad\mathrm{with}\, \epsilon'_{\alpha,\lambda}\equiv\frac{d\epsilon_{\alpha,
\lambda}}{d\lambda}\,\textrm{ and } p'_{\alpha,\lambda}=\frac{d p_{\alpha,\lambda}}{d\lambda} , 
\end{equation}
where $\epsilon_{\alpha,\lambda}$ is the bare energy of a quasiparticle defined in~\eqref{xxz-en}, and $p_{\alpha,\lambda}$ its bare momentum with $p'_{\alpha,\lambda} = 2 \pi a_{\alpha,\lambda}$. 

Interestingly, for generic integrable models the 
quasiparticle velocities depend on the thermodynamic macrostate~\cite{BEL:lightcone}, 
and  they are ``dressed'' by the interactions. This happens because in  interacting integrable models 
the addition or removal of a single quasiparticle causes a global 
rearrangement of the rapidities of the other quasiparticles. The net effect 
is a ``dressing'' of the bare quasiparticles properties, including 
the energies $\epsilon_{\alpha,\lambda}$, and hence the group velocities. 

The correspondence between thermodynamic macrostates and microscopic 
eigenstates of~\eqref{xxz-ham} is not one-to-one. 
In fact, the densities $\rho_{\alpha,\lambda}$ and $\rho_{\alpha,
\lambda}^{\scriptscriptstyle(\mathrm{h})}$ do not uniquely 
determine a microscopic eigenstate. In the thermodynamic limit, the number of microscopic eigenstates that give 
  rise to the same set of macroscopic densities diverges exponentially with the system size. 
The number of these thermodynamically equivalent eigenstates is 
given in terms of the so-called Yang--Yang entropy \cite{Taka99} as 
\begin{equation}
\frac{1}{L}\,\ln(\textrm{\# of eigenstates}) = s_{\mathrm YY} = \sum_{\alpha=1}^{\infty} 
\int_{-\pi/2}^{\pi/2} d\lambda s_{\alpha,\lambda},
\label{eq:YYEntropy}
\end{equation}
with the entropy density function $s_{\alpha,\lambda}$ being
\begin{equation}
	s_{\alpha,\lambda}=\rho_{\alpha,\lambda}[\log(1 + \eta_{\alpha,\lambda}) + \eta_{\alpha,\lambda} \log (1 + \eta_{\alpha,\lambda}^{-1})].
\label{eq:s}
\end{equation}

The TBA formalism has been applied  to describe thermal properties 
of the XXZ chain~\cite{Taka99}. The corresponding thermodynamic ensemble 
is the Gibbs ensemble, and  the Yang-Yang entropy~\eqref{eq:YYEntropy} 
is the usual thermal entropy. However, the TBA framework can also be used to describe 
the thermodynamic macrostate arising after a quantum quench,
i.e., the macrostate described by a generalized Gibbs ensemble (GGE) that takes into account the conservation of all the quasilocal charges. 
Then, the corresponding Yang-Yang entropy becomes the GGE thermodynamic entropy. 
Remarkably, this Yang-Yang entropy coincides with the von Neumann 
  entanglement entropy of the post-quench steady state~\cite{kauf,dls-13,collura-2014,nahum-17,nwfs-18,AlCa17}, 
and it is one of the main ingredients to reconstruct the full-time 
dynamics of the entanglement entropy (as it is clear from~\eqref{semi-i}).

\subsection{TBA treatment of the steady state after quenches from homogeneous states}
\label{sec-tba-quench}

Integrable models possess an 
extensive number of local and quasilocal conserved quantities. Their expectation value 
in the initial state is preserved during the dynamics. 
Thus, the post-quench dynamics is strongly constrained, implying that 
a Generalized Gibbs Ensemble, instead of the standard Gibbs one, has to be used to describe 
local properties of the steady state  in the long time limit. The GGE can be thought of as emerging 
from a generalized microcanonical 
principle~\cite{ggemc}. The eigenstates entering in the microcanonical average are 
the ones that have the correct expectation value of the local and 
quasilocal conserved quantities. In the thermodynamic limit the 
vast majority of these eigenstates 
give rise to the same set of densities $\rho_{\alpha,\lambda}$. This set of densities is called the representative state.  
The corresponding hole densities $\rho^{\scriptscriptstyle(h)}_{\alpha,\lambda}$ are obtained 
from the BGT equations~\eqref{eq:BGT}. The representative state encodes all information 
about local properties of the steady state. 

Within the TBA approach there are several techniques to determine the densities $\rho_{\alpha,\lambda}$ that describe the representative state.  
For instance, they can be determined from the overlaps between the eigenstates of the 
XXZ chain with the initial state, by using the 
so-called Quench Action method~\cite{CQA16}. Alternatively, they can be calculated 
from the knowledge of the initial values of the local and quasilocal conserved 
quantities~\cite{string-charge}. The latter method allows to deal, in 
principle, with any translationally invariant initial state. 

It is customary to describe the representative state in terms of $\eta_{\alpha,\lambda}$. These functions satisfy the so-called 
Y-system~\cite{suzuki-99}, which leads to the set of recursive 
equations for the functions $\eta_{\alpha,\lambda}$ (cf.~\eqref{fill-eta}) 
as 
\begin{equation}
\eta_{\alpha,\lambda} = 
\frac{\eta_{\alpha-1,\lambda+i\eta/2}\eta_{\alpha-1,\lambda-i\eta/2}
}{1 + \eta_{\alpha-2,\lambda}}-1\qquad\mathrm{with}\,\alpha\ge 2,
\label{equation:Ysystem}
\end{equation}
with the convention that $\eta_{0,\lambda}\equiv 0$. Once $\eta_{1,\lambda}$ is known, then the functions $\eta_{\alpha,\lambda}$ for $\alpha>1$ can be computed using \eqref{equation:Ysystem}. 
The corresponding particle densities $\rho_{\alpha,\lambda}$ can be 
computed by substituting the $\eta_{\alpha,\lambda}$ in the BGT 
equations~\eqref{eq:BGT}. 

Clearly, Eq.~\eqref{equation:Ysystem} implies that to determine the steady-state 
properties after a homogeneous global quench, one needs to calculate only 
$\eta_{1,\lambda}$. For both the tilted N\'eel state and the Majumdar-Ghosh state, 
which are relevant for this work, the function $\eta_{1,\lambda}$ is exactly known. 
For the tilted N\'eel state $|\mathrm{N},\theta\rangle$ with tilting 
angle $\theta$, one has~\cite{piroli-2016} 
\begin{equation}
1 + \eta_{1,\lambda} = \frac{T_{1}(\lambda+
i\frac{\eta}{2})}{\phi(\lambda+i\frac{\eta}{2})}
\frac{T_{1}(\lambda-i\frac{\eta}{2})}
{\bar\phi(\lambda-i\frac{\eta}{2})},
\label{eq:eta1TiltedNeel}
\end{equation}
where
\begin{eqnarray}
T_{1}(\lambda) &= -\frac{1}{8}\cot(\lambda)\{8\cosh(\eta)
\sin^{2}(\theta)\sin^{2}(\lambda)-4\cosh(2\eta) \\&
\quad+[\cos(2\theta)+3][2\cos(2\lambda)-1] + 
2\sin^{2}(\theta)\cos(4\lambda)\},
\label{eq:T1}
\end{eqnarray}
and
\begin{eqnarray}
\phi(\lambda) &= \frac{1}{8} \sin(2\lambda+i\eta)[2
\sin^{2}(\theta)\cos(2\lambda-i\eta)+\cos(2\eta)+3],
\label{eq:phi} \\
\bar\phi(\lambda) &= \frac{1}{8} \sin(2\lambda-
i\eta)[2\sin^{2}(\theta)\cos(2\lambda+i\eta)+\cos(2\eta)+3].
\label{eq:phibar}
\end{eqnarray}
For the dimer state, the funcion $\eta_{1,\lambda}$ 
reads~ \cite{string-charge}
\begin{equation}
\eta_{1,\lambda} = \frac{\cos(4
\lambda)-\cosh(2\eta)}{\cos^{2}(\lambda)(\cos(2\lambda)-\cosh(2\eta))}-1.
\label{eq:eta1Dimer}
\end{equation}
%

\subsection{Quenches from piecewise homogeneous initial states: Generalized Hydrodynamics}
\label{ghd}

Here we consider quenches from piecewise homogeneous initial states (as described in Fig.~\ref{fig0} (b)). 
Two semi-infinite chains  $L$ and $R$ are prepared in the  translationally invariant states $|\Psi_L\rangle$ and 
$|\Psi_R\rangle$, and are suddenly joined together at $t=0$. The ensuing dynamics 
  is governed by the globally translational invariant Hamiltonian \eqref{xxz-ham}. 
Recently, it has been shown that a Generalized Hydrodynamics (GHD) approach 
allows to study this quench~\cite{CaDY16,BCDF16} in the long time limit at large spatial scales. 
Physically, during time evolution a  light-cone spreads from 
the interface between the two chains. 
Outside of this lightcone, the properties of the system are the 
same as after the homogeneous quenches from the states 
$|\Psi_L\rangle$ and $|\Psi_R\rangle$ (see Fig.~\ref{fig0} (a)). 
Inside the lightcone and at late times, the expectation values of local and quasilocal 
  observables become  functions of $\zeta=x/t$. This 
reflects the propagation of stable quasiparticles between the two 
chains. This also suggests the emergence of a local quasi-stationary 
state for each $\zeta$. For integrable models, this corresponds to a 
$\zeta$-dependent GGE. Within the TBA framework, 
the GGE is represented by a set of TBA densities 
$\{\rho_{\alpha,\lambda}(\zeta)\}_{\alpha=1}^\infty$. 

The key result of the GHD is that because of infinitely many conserved charges, 
the densities $\rho_{\alpha,\lambda}(\zeta)$ satisfy a simple 
continuity equation~\cite{CaDY16,BCDF16} as 
\begin{equation}
\label{ghd-eq}
 \partial_t\rho_{\alpha,\lambda}(\zeta)+\partial_x (v_{\alpha,\lambda}(\zeta)
\rho_{\alpha,\lambda}(\zeta))=0. 
\end{equation}
In the bipartite quench the dependence of local observables on the space-time 
coordinates $x,t$ is only through $\zeta$. 
Here $v_{\alpha,\lambda}(\zeta)$ are the dressed 
group velocities (the same as in~\eqref{semi-i}), 
which are solutions of the system of 
integral equations~\cite{BEL:lightcone}
\begin{equation}
  v_{\alpha,\lambda}(\zeta) = v_{\alpha,\lambda}^{\textrm{\tiny bare}}
  (\zeta) + \sum_\beta\int_{-\pi/2}^{\pi/2} d\mu \frac{T_{\alpha\beta}
(\lambda-\mu)}{a_{\alpha,\lambda}} \rho_{\beta,\mu}{ (\zeta)} 
{ (v_{\alpha,\lambda}(\zeta)-v_{\beta,\mu}(\zeta))}, 
\label{eq:vDoyon}
\end{equation}
where $T_{\alpha\beta}$ is the scattering matrix (cf.~\eqref{eq:anm} for the 
result for the XXZ chain) and $a_{\alpha,\lambda}$ is defined in~\eqref{eq:an}. 
The functions $v_{\alpha,\lambda}^{\scriptscriptstyle\mathrm{bare}}$ are the 
bare velocities defined in~\eqref{v-bare}.

Physically, Eq.~\eqref{eq:vDoyon} reflects that, due to integrability, 
the scattering between the quasiparticles is elastic, and the 
only effect of the interactions is 
to renormalize the quasiparticles  velocities. Indeed, 
the term $\rho_{\beta,\mu} |v_{\alpha,\lambda}-v_{\beta,\mu}|$  in~\eqref{eq:vDoyon} 
is the number of quasiparticles with rapidity $\mu$ and of species 
$\beta$ that scatter in the unit time with the quasiparticle 
of species $\alpha$ and rapidity $\lambda$.  
The ratio $T_{\alpha\beta}(\lambda-\mu)/a_{\alpha,\lambda}$ can be 
interpreted as an effective shift of the trajectory of the quasiparticle 
with label $\alpha,\lambda$ due to the scatterings. 
This interpretation underlies the flea gas method (cf.~section~\ref{sec-flea}). 

The GHD approach 
has been successfully applied to describe transport properties in 
spin systems, one-dimensional integrable field 
theories, both classical and quantum~\cite{DDKY17b,BePi17,Bulchandani-17,DoYo17,DoYo17a,IlDe17,BVKM17,cdv-17,BePC18,PDCB17,DSY17,MBPC18,mvc-18,bertini-2019,spohn-2019,doyon-2019,myers-2019,alvise-2018,vu-2018,mazza-2018}.  
Very recently, it has been shown that GHD provides a precise 
framework to describe experiments with trapped cold 
atoms~\cite{jerome-2018}. A recent interesting direction is to 
generalize the approach to include diffusive 
corrections~\cite{null,DBD,de-nardis-2018a,GHKV,sarang-2018a}. 
Finally, the GHD approach can be used to study the entanglement dynamics 
after quenches from piecewise-homogeneous initial states~\cite{Alba17,alba-2018,alba-2019}.

Unfortunately, calculating the full-time entanglement dynamics is 
in general a demanding task. The reason is that inside the lightcone 
(see Fig.~\ref{fig0} (b)) the trajectories of the quasiparticles 
are not straight lines. Explicit results are easily obtained only in 
some regimes. For instance, the steady-state value of the von Neumann 
entropy of a finite interval placed 
next to the interface between the two chains~\cite{Alba17}, as well as 
the growth rate of the entanglement entropy between two-semi-infinite 
systems~\cite{alba-2019}, can be calculated in terms of the $\zeta=0$ macrostate only. 

\section{Flea gas approach for out-of-equilibrium integrable systems}
\label{sec-flea}

The flea gas  approach was introduced in Ref.~\cite{DDKY17a} as an effective numerical 
method to simulate the GHD by employing classical ``molecular dynamics'' 
techniques. The method allows to simulate the dynamics of a quantum system 
starting from any thermodynamic macrostate, both homogeneous as well as 
inhomogeneous. So far, the approach has been implemented for the 
Lieb-Liniger gas but not for lattice systems such as the XXZ model. 

The method was inspired by the  correspondence between 
the continuity equation~\eqref{ghd-eq} and the hydrodynamic 
equations of a system of classical particles (hard-rod gas). 
Hard rods are classical one-dimensional objects undergoing 
elastic scattering. Here we denote their length as $d$. The hard 
rods dynamics is as follows. Hard rods move like free particles 
with bare velocity $v_\mathrm{b}$. When the distance between the centers 
of two hard rods equals $d$, they exchange their velocities. 
Following Ref.~\cite{DDKY17a}, here we adopt an alternative 
interpretation. One can think of hard 
rods as point-like objects. The scattering is then implemented as follows. 
When two hard rods are at the same point in space they scatter. The scattering 
consists of an instantaneous displacement by length $d$ 
of the positions of the two particles. 
Precisely, after assuming $d>0$ we impose that the particle on the left (right) 
is shifted by $d$ to the right (left). 


Let us define the density of rods with ``bare'' velocity $v_{\mathrm{b}}$ as 
$\rho(v_\mathrm{b})$. The number of rods with velocity between $v_{\mathrm{b}}$ and 
$v_{\mathrm{b}}+dv$ and in the spatial interval $dx$ is 
$\rho(v_{b}) dv dx$. During time evolution, many scatterings 
occur. The net effect is a renormalization of the velocity of the hard rods. 
Let us define this space-time dependent renormalized or ``dressed'' velocity as $v(v_\mathrm{b};x,t)$ 
The dressed velocity is a function of the bare velocity $v_\mathrm{b}$. 
The density $\rho(v_\mathrm{b})$ 
obeys the continuity equation~\cite{boldrighini-1983} 
\begin{equation}
\label{hr-c}
\partial_t\rho(v_\mathrm{b};x,t)+\partial_x(v(v_\mathrm{b};x,t)\rho(v_\mathrm{b};x,t))=0. 
\end{equation}
The renormalized velocity is given by the integral 
equation~\cite{boldrighini-1983} 
\begin{equation}
\label{hr-v}
v(v_\mathrm{b};x,t)=v_\mathrm{b}+d\int dw 
\rho(w;x,t)(v(v_\mathrm{b};x,t)-v(w;x,t)). 
\end{equation}
Equation~\eqref{hr-v} has the same structure as~\eqref{eq:vDoyon}, and it 
admits a simple interpretation. The expression 
$\rho(w)|v(v_\mathrm{b})-v(w)|$  is the number of 
scatterings per unit time between the hard rod with velocity $v$ and the ones with 
velocity $w$. 
The second term on the right hand side in~\eqref{hr-v} 
is the total shift that happens in the unit time to the trajectory 
of the hard rod with bare velocity $v$ due to the scatterings with 
other hard rods.

\subsection{Flea gas for the XXZ chain and numerical implementation}
\label{sec-flea-xxz}

We now discuss the application of the flea gas method for the XXZ chain. Before 
describing the method for the quenches from an inhomogeneous initial state, it is 
useful to consider the case of homogeneous ones. 
The TBA densities $\rho_{\alpha,\lambda}$  describing the steady state 
after the quench are stationary and homogeneous, i.e., they do not depend 
on $x,t$. The group velocity $v_{\alpha,\lambda}$ of the quasiparticles 
are obtained by solving the TBA system~\eqref{eq:vDoyon}. 
The crucial observation is that Eq.~\eqref{eq:vDoyon} has the same structure 
as the equation for the hard rod gas~\eqref{hr-v}. Equation~\eqref{eq:vDoyon} 
can be interpreted as the dressing equation for the velocities of a 
system of multi-species and point-like classical particles 
undergoing elastic scattering. Now each particle is identified by a double 
index $(\alpha,\lambda)$, and  $T_{\alpha,\beta}(\lambda-\mu)/a_{\alpha,\lambda}$ is 
 interpreted as a scattering length. Thus, we define 
$d_{\alpha,\beta}(\lambda,\mu)$ as 
\begin{equation}
\label{flea-d}
d_{\alpha,\beta}(\lambda,\mu)=\frac{T_{\alpha,\beta}
(\lambda-\mu)}{a_{\alpha,\lambda}}. 
\end{equation}
Similarly to the hard rods, the particles move freely with bare velocities 
$v^{\scriptscriptstyle\mathrm{bare}}$ (now given by~\eqref{v-bare}). Scatterings occur 
when two particles are at the same point in space. 
If the particle coming from the left has labels $(\alpha,\lambda)$, and the 
particle coming from the right has labels $(\beta,\mu)$, then
the particle coming from the left will jump $d_{\alpha,\beta}(\lambda,\mu)$,
and the particle coming from the right will jump $-d_{\beta,\alpha}(\mu,\lambda)$.

A crucial remark is in order. Unlike the case of the Lieb-Liniger model~\cite{DDKY17} 
it is not straightforward to show that the dynamics outlined above 
reproduces the correct dressing for the bare velocities of the particles, 
i.e., Eq.~\eqref{eq:vDoyon}. 
First, the displacement of the trajectory of a given particle is given as 
$\Delta x=v_{\alpha,\lambda}\Delta t$, which defines the dressed 
velocity $v_{\alpha,\lambda}$. The dressing of 
the velocity arises from the scattering with the other particles. 
The number of scatterings per unit time between a particle with label 
$(\alpha,\lambda)$ and particles with label $(\beta,\mu)$ is 
given as $\rho_{\beta,\mu}|v_{\alpha,\lambda}-
v_{\beta,\mu}|\Delta t$. 

The key issue is how to determine the direction of the jump. 
During the flea gas dynamics the particles move with their bare 
velocities. 
If a particle moving at  $v_{\alpha,\lambda}^{\scriptscriptstyle\mathrm{bare}}$ 
scatters with another one with  velocity $v_{\beta,\mu}^{\scriptscriptstyle\mathrm{bare}}$ 
its trajectory gets shifted by $\mathrm{sign}(v_{\alpha,\lambda}^{\scriptscriptstyle\mathrm{bare}}-
v_{\beta,\mu}^{\scriptscriptstyle\mathrm{bare}})d_{\alpha,\beta}(\lambda,\mu)$. 
For the Lieb-Liniger gas one can show that the dressed velocities are monotonic functions 
of the bare ones, which implies that $\mathrm{sign}(v_{\alpha,\lambda}^{\scriptscriptstyle\mathrm{bare}}- v_{\beta,\mu}^{\scriptscriptstyle\mathrm{bare}})=
\mathrm{sign}(v_{\alpha,\lambda}-v_{\beta,\mu})$. This ensures that the jumped 
length is $\mathrm{sign}(v_{\alpha,\lambda}-v_{\beta,\mu}) d_{\alpha,\beta}
(\lambda,\mu)$. By summing over $\beta$ and integrating over $\mu$, one obtains 
the term on the right-hand-side in~\eqref{eq:vDoyon}. This shows that the 
flea gas dynamics gives the correct dressing for the group velocities of 
the particles. On the other hand, for the XXZ chain the dressed 
velocities are not monotonically increasing functions of the bare ones. 
An important consequence is that now $\mathrm{sign}(v_{\alpha,\lambda}^{
\scriptscriptstyle\mathrm{bare}}- v_{\beta,\mu}^{\scriptscriptstyle
\mathrm{bare}})\ne\mathrm{sign}(v_{\alpha,\lambda}-v_{\beta,\mu})$. 
This implies that  for the XXZ chain one cannot conclude that the 
total jumped length is given as $(v_{\alpha,\lambda}-v_{\beta,\mu}) 
d_{\alpha,\beta}(\lambda,\mu)$. 
To overcome this problem, our strategy here is to use the flea gas dynamics as outlined above, 
showing numerically that, at least in the quenches that we consider, 
it gives the correct dressing for the group velocities of the 
particles (see section~\ref{sec-ent-h}). 
   
We now discuss the details of the implementation of the flea gas 
method for the XXZ chain.  The system is in the continuum, and it is of length $L$. 
Both space and time are treated as continuous variables. 
For a homogeneous quench, the initial state of the simulation is prepared as 
follows.  First, we create a total number of 
particles $N_p$. The particles are described by the TBA densities 
$\rho_{\alpha,\lambda}$, which contain the full information about the 
post-quench GGE (see section~\ref{sec-tba-quench} for the results for 
the quenches considered here). $N_p$ is chosen such that one has the 
correct value of the particle density, i.e., 
\begin{equation}
N_p=L\sum_\alpha\int d\lambda \rho_{\alpha,\lambda}
\end{equation}
Note that $N_p$ is not the total number of down spins $N$, which is given as 
$N=L\sum_\alpha\int d\lambda \alpha\rho_{\alpha,\lambda}$. This simply 
reflects that in the simulation multi-spin bound states are treated as 
individual point-like particles. The particles are labeled as $1,\dots,{N_p}$. 
Here we restrict ourselves to the situation in which only pairs of entangled quasiparticles 
with opposite rapidities~\cite{AlCa17} are present. For convenience, particles 
forming an entangled pair are labelled by consecutive integers 
as $(2\gamma,2\gamma+1)$ with  $\gamma=1,\dots,N_p/2$. 
To each pair we assign a species label $\alpha$ with probability 
$r_\alpha$ given as 
\begin{equation}
  r_\alpha=\frac{L}{N_{\mathrm p}} \int d\lambda \rho_{\alpha,\lambda}. 
\end{equation}
Similarly,  rapidities {$\lambda_{2\gamma}=-\lambda_{2\gamma+1}$} are assigned to the pairs 
with probability $\rho_{\alpha,\lambda}=\rho_{\alpha,-\lambda}$. 
 The position of each pair is random in the interval $[-L/2,L/2]$. Note that entangled particles are 
produced at the same point in space, implying {$x_{2\gamma}=x_{2\gamma+1}$}. 
However, to avoid spurious scatterings when the dynamics starts, we impose a tiny 
displacement between entangled particles. 
Finally, we assign to each pair their contribution to the Yang-Yang entropy, which is $s_{\alpha,\lambda}/\rho_{\alpha,\lambda}$ 
(cf.~\eqref{eq:s}). 

During the  time evolution, the particles move with the bare 
velocities $v^{\scriptscriptstyle \textrm{\tiny bare}}_{\alpha,\lambda}$, 
given as (cf~\eqref{v-bare}) 
\begin{equation}
v_{\alpha,\lambda}^{\textrm{\tiny bare}}=
 \frac{\sinh(\eta)\,a'_{\alpha,\lambda}}
{2 a_{\alpha,\lambda}}.
 \label{eq:vBare}
\end{equation}
Here $a_{\alpha,\lambda}$ are defined in~\eqref{eq:an} and $a'_{\alpha,\lambda}
\equiv da_{\alpha,\lambda}/d\lambda$. 
During the simulation only the position of the particles are 
updated, whereas their labels, velocities, and entropies remain the same. 
Particles can collide, jumping backward and forward of distance 
$d_{\alpha,\beta}(\lambda-\mu)$ (cf.~\eqref{flea-d}). 
This happens as follows. Let us denote two colliding particles as 
$P_1$ and $P_2$, $P_1$ being the left particle and $P_2$  
the right one, respectively. Let us assume that $P_1$ and $P_2$ 
have labels $(\alpha,\lambda)$ and  $(\alpha',\lambda')$, respectively. 
 Thus, $P_1$ jumps to the right of distance 
$d_{\alpha,\alpha'}(\lambda,\lambda')$, whereas 
particle $P_2$ jumps to the left of distance 
$d_{\alpha',\alpha}(\lambda',\lambda)=-d_{\alpha,\alpha'}(\lambda,\lambda')$. 
It is crucial to observe that while jumping, $P_1$ and $P_2$ can scatter 
with other particles that are within $d_{\alpha,\alpha'}(\lambda,\lambda')$. 
For example, if the trajectory of $P_1$, after scattering with 
$P_2$, crosses that of a third particle $P_3$ with label $(\alpha'',\lambda'')$, 
$P_1$ scatters with $P_3$, as well. This means that, in principle,  
there is a ``cascade'' of scatterings initiating when $P_1$ and $P_2$ 
collide. 

The complete flea gas algorithm  is illustrated in Fig.~\ref{algo-flea}. 
The first step is to identify the pair of particles $P_1$ and $P_2$ 
that scatter first, and the corresponding scattering time $t_{\mathrm{\mathrm{coll}}}$. 
This is performed by the routine $\textsc{Find}(P_1,P_2,t_\mathrm{coll})$ 
in Fig.~\ref{algo-flea}. This can be done efficiently by using standard methods in 
molecular dynamics simulations (see for instance Ref.~\cite{smac}). 
Then, all the particles are evolved until 
$t_{\mathrm{coll}}$, when the scattering between $P_1$ and $P_2$ occurs. 
This is described by the procedure \textsc{Collide} 
in Fig.~\ref{algo-flea}. $P_1$ and $P_2$ are instantaneously displaced by a 
distance $d_{1,2}$ and $d_{2,1}$ (cf.~\eqref{flea-d}). 
The displacement of the particles is 
implemented with the procedures \textsc{JumpLeft} and \textsc{JumpRight}, 
which are described in Fig.~\ref{alg-flea-jumps}. 
Note that the two scattering particles are marked before starting the 
collision (see procedure \textsc{Mark}). 
This is to prevent that, 
while  scattering with near particles, $P_1$ and $P_2$ scatter again with 
each other.  Marked particles, instead of scattering, cross each 
other. After the scattering cascade starting with their first collision happened, 
$P_1$ and $P_2$ are unmarked. 
\begin{algorithm}[t]
\begin{multicols}{2}
\begin{algorithmic}[1]
\Procedure{Evolve}{$t_{\mathrm{max}}$}
\State $t=0$
\While{$t<t_{\mathrm{max}}$}
\State \textsc{Find}($P_1,P_2,t_{\mathrm{coll}}$)
\State $\forall\gamma,\quad x_\gamma\rightarrow x_\gamma+v_\gamma t_{\mathrm{coll}}$
\State $\forall\gamma$, \textsc{Unmark}($P_\gamma$)
\State \Call{Collide}{$P_1,P_2$}
\State $t=t+t_{\mathrm{coll}}$
\EndWhile
\EndProcedure
\columnbreak
\Procedure{Collide}{$P_1,P_2$}
\If{\textsc{Marked}($P_1,P_2$)}
\State $x_1\leftrightarrow x_2$
\Else
\State \textsc{Mark}($P_1,P_2$)
\State $x_1\leftrightarrow x_2$
\State \Call{JumpRight}{$P_1,d_{12}$}
\State \Call{JumpLeft}{$P_2,d_{21}$}
\EndIf
\EndProcedure
\end{algorithmic}
\end{multicols}
\hrule
\vspace{-.2cm}
\captionof{figure}{Flea gas dynamics. The main procedure \textsc{Evolve} 
 evolves the system up to $t_{\mathrm{max}}$. The routine \textsc{Find}($P_1,P_2,t_{\mathrm{coll}}$) 
 finds the particles $P_1$ and $P_2$ that 
 scatter first at time $t_{\mathrm{coll}}$. The 
 positions $x_\gamma$ of the particles are evolved 
 up to $t_\mathrm{coll}$. $v_\gamma$ are the bare velocities (cf.~Eq.~\eqref{eq:vBare}). 
 Then, particles $P_1$ and $P_2$ scatter. The function $\textsc{Unmark}$ 
 removes the mark assigned to the particles when they scatter for the first time. The 
 scattering is  implemented by \textsc{Collide}:  $P_1$ and $P_2$ are 
 displaced by a distance $d_{12}=-d_{21}$ 
 (cf.~Eq.~\eqref{flea-d}). 
 The functions \textsc{JumpLeft} and \textsc{JumpRight} 
 implementing this shift are in Fig.~\ref{alg-flea-jumps}. 
 Before scattering the particles are marked. Marked particles 
 cross each other. Note that while $P_1$  is scattering 
 with $P_2$, a scattering with a third particle $P_3$ can occur, initiating 
 a scattering ``cascade''. 
}
\label{algo-flea}
\end{algorithm}

\begin{algorithm}[t]
\begin{multicols}{2}
\begin{algorithmic}[1]
\Procedure{JumpRight}{$P_1,d$}
\While{$d>0$}
\If{$|x_3-x_1|>d$}
\State $x_1=x_1+d$
\State $d=0$
\Else
\State $d=d-|x_3-x_1|$
\State $x_1=x_3$
\State \Call{Collide}{$P_1,P_3$}
\EndIf
\EndWhile
\EndProcedure

\Procedure{JumpLeft}{$P_1,d$}
\While{$d<0$}
\If{$|x_1-x_3|<|d|$}
\State $x_1=x_1+d$
\State $d=0$
\Else
\State $d=d+|x_1-x_3|$
\State $x_1=x_3$
\State \Call{Collide}{$P_3,P_1$}
\EndIf
\EndWhile
\EndProcedure
\end{algorithmic}
\end{multicols}
\hrule
\vspace{-.2cm}
\captionof{figure}{Jump algorithms for the flea gas dynamics. When 
 scattering with each other, particles $P_1$ and $P_2$ are 
 instantaneously displaced by a distance $d$, which depends 
 on the species and the rapidity of the particles, and it is extracted from the 
 scattering matrix of the model (see Eq.~\eqref{flea-d}). The functions \textsc{JumpRight} 
 and \textsc{JumpLeft} implement this displacement. In \textsc{JumpRight} and \textsc{JumpLeft} 
 particle $P_3$ is the next particle on the right and on the left of $P_1$, respectively. 
 If during the jump $P_1$ does not meet particle $P_3$, the position of $P_1$ is shifted by $d$. 
 If $|x_1-x_3|<|d|$, then particle $P_1$ scatters with $P_3$. 
 The procedure \textsc{Collide} is defined in Fig.~\ref{algo-flea}. 
}
\label{alg-flea-jumps}
\end{algorithm}

\subsection{Entanglement dynamics in flea gas simulations}
\label{sec-flea-ent}

The entanglement entropy at a given time is computed by 
counting the entangled pairs (weighted with their Yang-Yang entropy) 
that are shared between the 
subsystem of interest $A$ (cf.~Fig.~\ref{fig0}) and the rest, i.e.,  the number of 
pairs $(P_\gamma,P_{\gamma+1})$, such that $x_{\gamma}$ and 
$x_{\gamma+1}$ are in different subsystems. The result  
for the entanglement entropy reads as 
\begin{equation}
S(t) =  \Big\langle\sum_{\textrm{\tiny shared 
pairs $(\lambda,-\lambda)$}}  \frac{s_{\alpha,\lambda}}{\rho_{\alpha,\lambda}} 
\Big\rangle_{t}. 
\label{eq:entropyFleaGas}
\end{equation}
Here the average $\langle\rangle_t$ is over different 
realizations of the flea gas dynamics up to time $t$. 
The sum is over the pairs that are shared between the two subsystems. 
Importantly, the factor $1/\rho_{\alpha,\lambda}$ takes into account 
that different types $(\alpha,\lambda)$ of particles appear in the
sum~\eqref{eq:entropyFleaGas} with a frequency $\ell \rho_{\alpha,\lambda}$. 

\section{Numerical results}
\label{sec-ent-num}

We now provide numerical results showing the validity of the flea gas 
method to calculate the dynamics of the entanglement entropy after a 
generic quench in integrable systems. In section~\ref{sec-ent-h} 
we present some preliminary benchmarks of the approach. 
In section~\ref{sec-ent-in} we discuss the bipartite inhomogeneous quench 
depicted in Fig.~\ref{fig0} (b). Finally, in 
section~\ref{sec-ent-mi} we discuss the mutual information between two intervals.

\subsection{Preliminary benchmarks}
\label{sec-ent-h}

A crucial feature of the flea gas dynamics is that it gives rise 
to the correct dressing of the group velocities of the quasiparticles 
given by~\eqref{eq:vDoyon}. 
While this can be proven for the flea gas algorithm for the Lieb-Liniger 
model, this is not the case for the XXZ chain. 
Here we provide numerical evidence that, 
at least for the quenches that we consider, the flea gas dynamics 
presented in Section~\ref{sec-flea} gives rise to the correct 
dressing of the group velocities. 

\begin{figure}[t]
\begin{center}
\includegraphics*[width=1\linewidth]{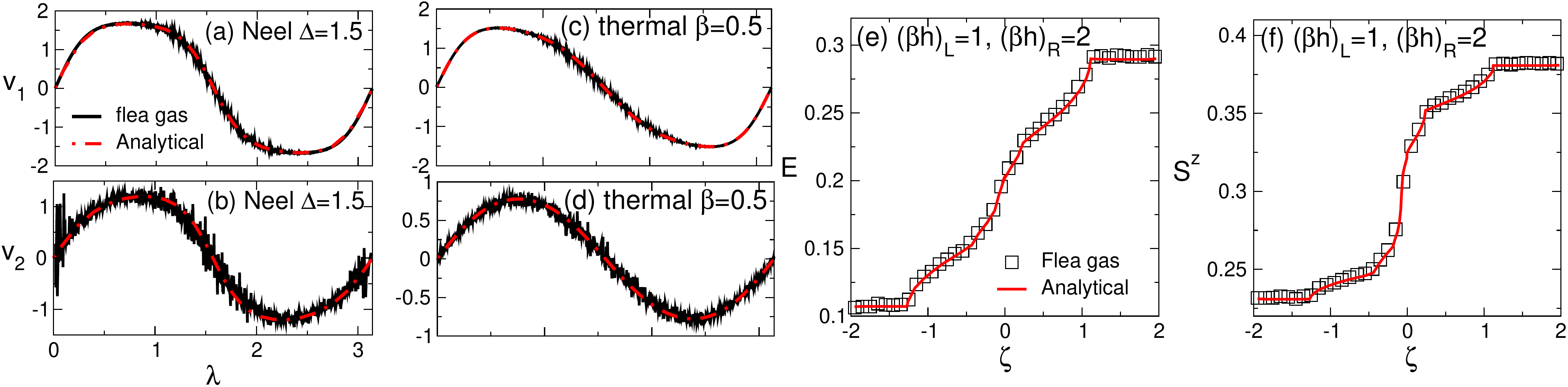}
\end{center}
\caption{ Flea gas versus GHD results.  Panels (a) and (b) show the 
 dressed group velocities $v_{\alpha,\lambda}$ for the first two strings 
 plotted versus the quasiparticle rapidity $\lambda$. 
 The black line is the flea gas result. The data are for a 
 chain with $L=2000$ sites and are averaged over $10^3$
 realizations of the dynamics. The dashed-dotted line is 
 the solution of~\eqref{eq:vDoyon}. 
 The results shown are for the quench from the N\'eel state and $\Delta=1.5$. 
 Panels (c) and (d) show the values of $v_1$ and $v_2$ for the 
 quench from the initial thermal density matrix~\eqref{thermal}. 
 Panels (e) and (f) show profiles of observables in the quench
 of the  XXZ chain with $\Delta=2$ from 
 the \emph{bipartite thermal} state with $\beta_{\mathrm{L}}=
 \beta_{\mathrm{R}}=0$, $(\beta h)_{\mathrm{L}}=1$, $(\beta 
 h)_{\mathrm{R}}=2$, considered in Ref. \cite{PDCB17}. 
 We plot the local energy $E$ and the magnetization 
 $S^z$ as a function of $\zeta\equiv x/t$. The 
 squares represent the flea gas results averaged over $10^3$ 
 realizations of the dynamics. The full line was obtained in \cite{PDCB17} by solving the GHD equations~\eqref{ghd-eq}. 
}
\label{fig:check}
\end{figure}

We first consider the quench from a homogeneous chain prepared in the 
N\'eel state. 
Our results are presented in 
Fig.~\ref{fig:check} (a) and (b). The results are for the XXZ 
chain with $\Delta=1.5$. The figures show the dressed 
velocities of the first two strings $v_{\alpha,\lambda}$ with 
$\alpha=1,2$ plotted versus the rapidity $\lambda$. The full lines 
are the flea gas results. These are obtained as 
$\Delta x_\gamma/t$, where $\Delta x_\gamma$ is the 
displacement of the particles with respect to their initial 
position. The data are averaged over $10^4$ realizations of the 
flea gas dynamics. As it is clear from the Figure, there are large 
fluctuations in the central region around $\lambda=\pi/2$. This is because 
the density $\rho_{1,\lambda}$ is large at the edges of the interval, 
whereas it is suppressed at the center, for instance, for $\Delta=1.5$ 
by a factor of $\sim 50$. This effectively reduces the statistics 
for the central rapidities. For $\alpha=2$ the density $\rho_{2,\lambda}$ 
has a maximum around $\lambda=\pi/2$. However, it is in general much 
smaller than $\rho_{1,\lambda}$, again resulting in large fluctuations 
for the group velocities of the two strings. In Fig.~\ref{fig:check} the dashed-dotted lines 
are the analytical results for the dressed velocities, which are 
obtained by solving numerically~\eqref{eq:vDoyon}. Clearly, 
the agreement with the flea gas results is very good. 
%
%

%
\begin{figure}[t]
\begin{center}
\includegraphics*[width=.55\linewidth]{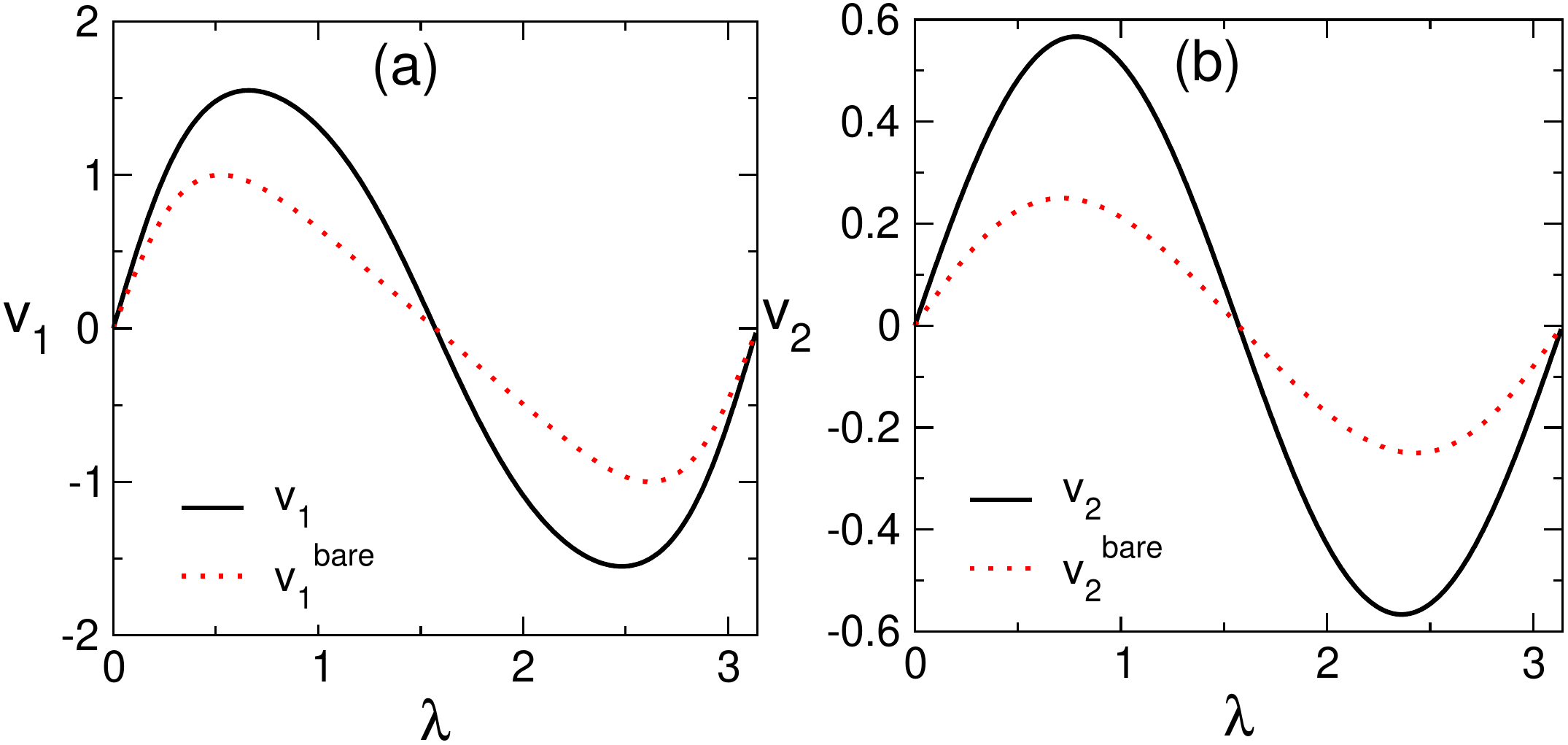}
\end{center}
\caption{ Comparison between bare and dressed velocities. The results 
 are for the thermal density matrix~\eqref{thermal} with $\beta=0.5$, 
 $\beta h=0.25$ and $\Delta=1.5$. The continuous line and the dotted 
 line are the dressed and the bare velocities, respectively. Panel (a) and 
 (b) show the velocities for the unbound and the two-particle bound 
 states. Note that non monotonicity of the velocities implies that 
 for some $\alpha,\beta,\lambda,\mu$ one has 
 $\mathrm{sign}(v_{\alpha,\lambda}^{\scriptscriptstyle\mathrm{bare}}
 -v_{\beta,\mu}^{\scriptscriptstyle\mathrm{bare}})\ne \mathrm{sign}(v_{\alpha,\lambda}-
 v_{\beta,\mu})$. 
}
\label{fig:vel}
\end{figure}
%

%
\begin{figure}[t]
\begin{center}
\includegraphics*[width=1\linewidth]{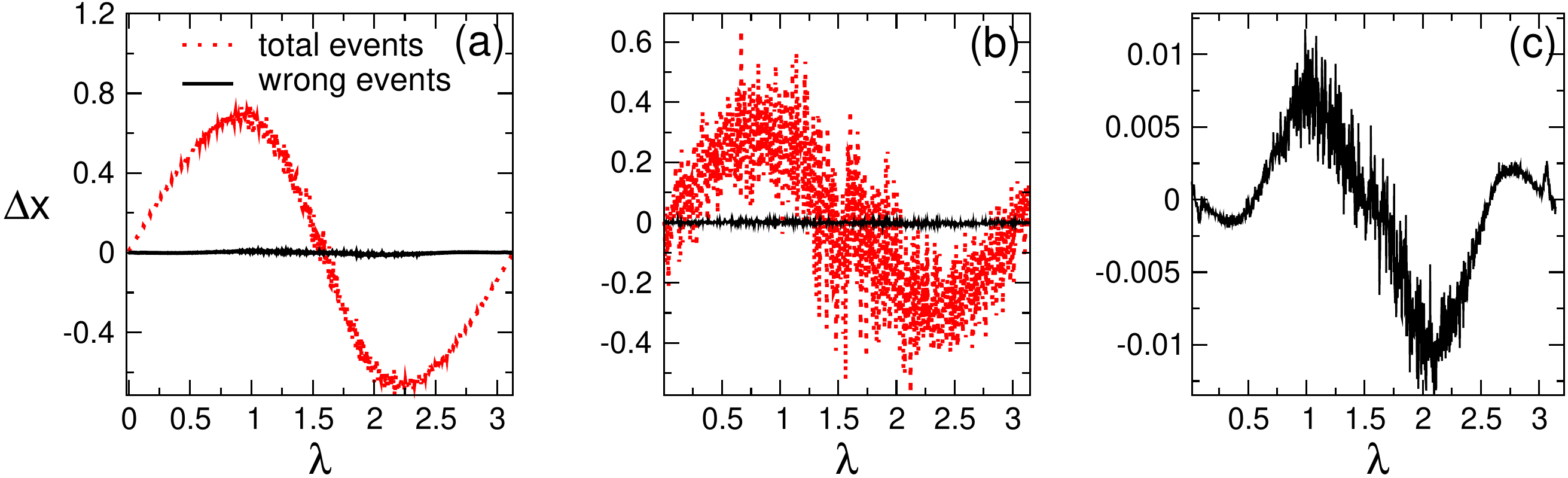}
\end{center}
\caption{ Effect of the ``wrong'' scattering events on the quasiparticles 
 trajectories. We show the shift $\Delta x$ experienced by the quasiparticles 
 with rapidity $\lambda$ due to scatterings with other particles. The data are for 
 the initial thermal density matrix in~\eqref{thermal} with $\beta=0.5$ and 
 $\beta h=0.25$. (a) 
 shows $\Delta x$ due to all the scatterings (dotted line) and only to 
 the wrong scattering events (continuous line) for the unbound particles. 
 In (b) we show the result for the two-particle bound states. Panel (c) shows 
 the shift due to the wrong collisions for the unbound particles (same data as 
 in (a)). 
}
\label{fig:coll}
\end{figure}
%

We also considered the dressed velocities in homogeneous thermal states. Panels (c) and (d) show results for the dressed velocities in the state described by the 
thermal density matrix 
\begin{equation}
	\label{thermal}
	\rho_0=\frac{1}{Z}e^{-\beta H+(\beta h) S^z},
\end{equation}
where $\beta$ is the inverse temperature and $h$ a transverse 
magnetic field. The data are for $\beta=0.5$ and $\beta h=0.25$. 
The continuous lines are flea gas results for the dressed 
velocities of the first two strings, which perfectly match 
the analytical results of TBA (dashed-dotted lines).

We now move the quenches from piecewise homogeneous states. Here we consider the 
initial density matrix as 
\begin{equation}
\label{thermal2}
\rho_0=\frac{1}{Z}e^{-\beta_L H_L+(\beta_L h_L)S^z_L}\otimes
e^{-\beta_R H_R+(\beta_R h_R)S^z_R}, 
\end{equation}
where quantities with the subscript $L/R$ refer to the left and right 
chains (see Fig.~\ref{fig0} (b)). The quench from~\eqref{thermal2} was 
investigated in Ref.~\cite{PDCB17} using GHD. Here we consider 
 $\beta_{\mathrm{L}}=0,\, (\beta h)_{\mathrm{L}}=1$, and  
 $\beta_{\mathrm{R}}=0,\,(\beta h)_{\mathrm{R}}=2$, with $h$ being the magnetic 
 field ($\Delta=2$). Due to the inhomogeneous initial condition now the dressed 
velocities depend on $\zeta\equiv x/t$ (see Fig.~\ref{fig0} (b)).  
To check the validity of the flea gas method, in principle, one has to 
check that the flea gas gives the correct result for $v_{\alpha,\lambda}(\zeta)$ 
for any $\zeta$. Here, instead, we consider the space-time dependence of 
the local energy density  $E(\zeta)$ and magnetization $S^z(\zeta)$ plotted 
versus $\zeta=x/t$, with $t$ the time after the quench, and $x$ the distance from 
the origin of the lightcone. Both the quantities for $x,t\to\infty$ become 
functions of $\zeta$. In Figure~\ref{fig:check} (e) and (f) 
the square symbols are the results of the flea gas simulation for a chain with 
$L=2000$ and $t=100$, whereas the full 
lines are the analytical results obtained 
in Ref.~\cite{PDCB17} by  solving the GHD equations. 
The agreement between the flea gas and the GHD results is spectacular. 

\begin{figure}[t]
\begin{center}
\includegraphics*[width=0.65\linewidth]{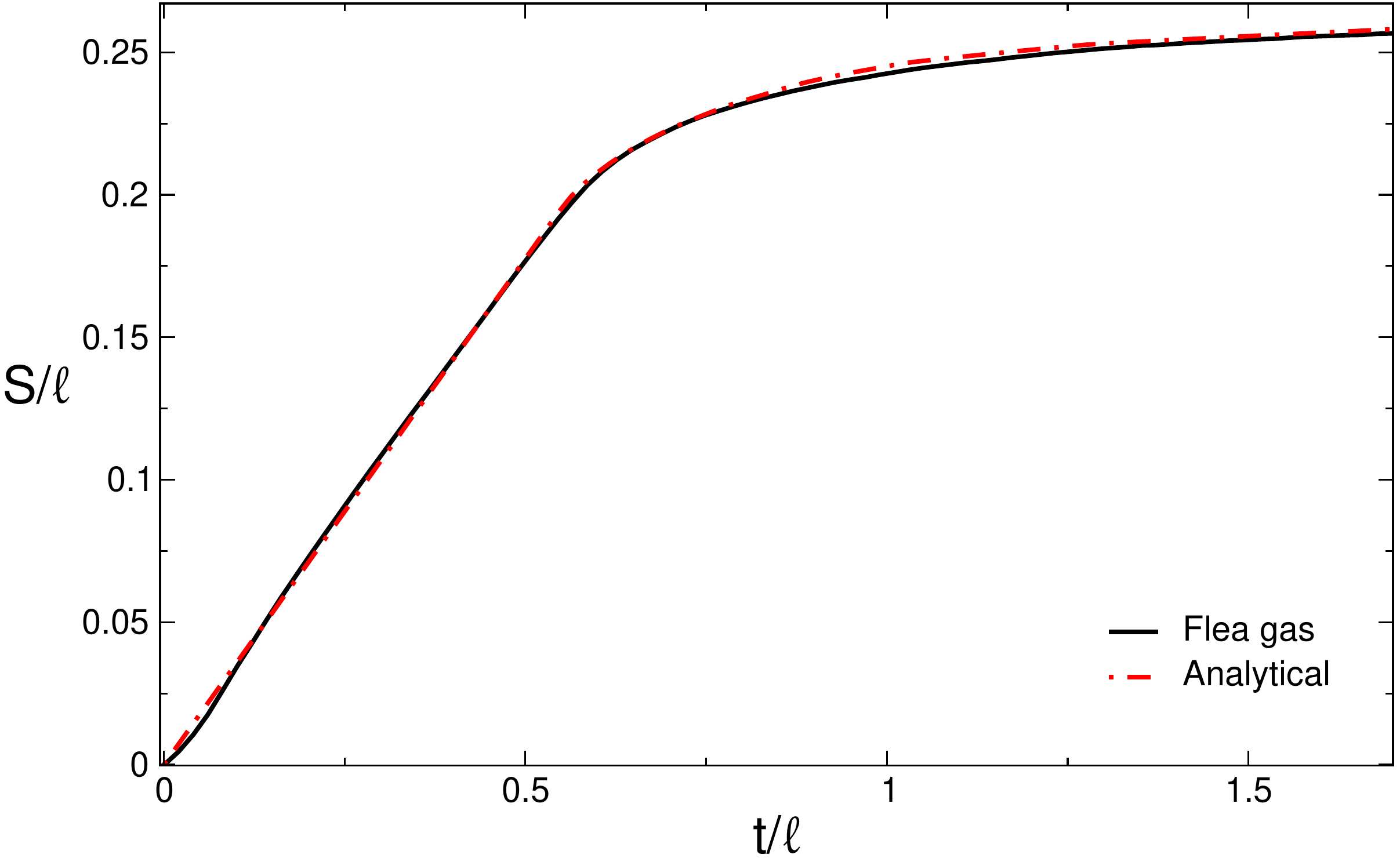}
\end{center}
\caption{ Dynamics of the von Neumann entanglement entropy after the 
 quench from the N\'eel state in the XXZ chain with 
 $\Delta=2$. The entropy density $S/\ell$ is plotted versus 
 the rescaled time $t/\ell$, with $\ell$ the subsystem size. 
 The continuous line corresponds to the flea gas simulation for a subsystem 
 with $\ell=100$. The length of the chain is $L=2000$. 
 The results are averaged over $\sim 1000$ 
 realization of the dynamics. The dashed-dotted line is the 
 analytical result~\eqref{semi-i}.}
\label{fig1}
\end{figure}

As a further check of the validity of the flea gas method we now 
discuss results for the dynamics of the von Neumann entanglement entropy after 
a quench from  homogeneous initial states, for which analytical 
results (cf. Eq.~\eqref{semi-i}) are available. 
Our results are discussed in Fig.~\ref{fig1}. The 
figure shows data for the XXZ chain with 
$\Delta=2$, quenching form the N\'eel state (see 
section~\ref{sec-quenches}). The rescaled entropy 
$S/\ell$ is plotted versus $t/\ell$, with $\ell$ the subsystem size. In the 
simulation we considered $\ell=100$ and   
a chain of length $L=2000$. The data are obtained by averaging 
over $\sim 10^4$ independent realizations of the flea gas 
dynamics. The continuous line is the flea gas result~\eqref{eq:entropyFleaGas} 
up to $t/\ell\approx 1.5$, although results for larger times can be easily obtained. 
The dashed-dotted line is the analytical result~\eqref{semi-i} 
obtained in Ref.~\cite{AlCa17}. The agreement between the two is excellent. 

Some remarks are in order. First, the flea gas picture is expected to capture 
correctly only the ballistic part of the entanglement dynamics, i.e., the leading behavior in 
$t/\ell$.  Note that, however,  subleading corrections, for instance 
diffusive corrections as ${\mathcal O}(\sqrt{t})$, are generically expected in the 
entanglement dynamics. 
In the flea gas framework diffusive corrections arise because of the  
average over the different realizations of the initial state,  and are associated 
with the fluctuations of the particles trajectories. On the other hand, 
the diffusive corrections that are present in the flea gas are not 
expected to be the same as the quantum diffusive corrections of the XXZ chain. 
The origin of diffusion in interacting integrable models 
and in the flea gas have been under constant investigation in the last few 
years~\cite{DBD,de-nardis-2018a,GHKV,sarang-2018a,null}. 
Finally, as it is clear from Fig.~\ref{fig1}, subleading corrections 
are small. Only for very short times some deviations from~\eqref{semi-i} 
are present, which disappear in the scaling limit $t,\ell\to\infty$. 
%
%
\subsubsection{Effect of the wrong collisions}
\label{sec:wrong}

In section~\ref{sec-flea-xxz} we stressed that the fact that the group 
velocity of the quasiparticles are not monotonic increasing functions 
of the bare ones. This implies that $\mathrm{sign}(v_{\alpha,\lambda}-v_{\beta,\mu})
\ne \mathrm{sign}(v_{\alpha,\lambda}^{\scriptscriptstyle\mathrm{bare}}
-v_{\beta,\mu}^{\scriptscriptstyle\mathrm{bare}})$. As a consequence 
two quasiparticles colliding with velocities $v_{\alpha,\lambda}^{
\scriptscriptstyle\mathrm{bare}}$ and 
$v_{\alpha,\lambda}^{\scriptscriptstyle\mathrm{bare}}$ are shifted by 
the ``wrong'' distance 
$\mathrm{sign}(v_{\alpha,\lambda}^{\scriptscriptstyle\mathrm{bare}}
-v_{\beta,\mu}^{\scriptscriptstyle\mathrm{bare}}) d_{\alpha,\beta}(\lambda,\mu)$. 

Here we investigate this effect. We consider the initial state defined 
by the thermal density matrix~\eqref{thermal} with $\beta=0.5$ and 
$\beta h=0.25$. We consider the XXZ chain with $\Delta=1.5$. The reason 
is that for $\Delta\to 1$  the dressing of the quasiparticles 
velocities is larger, which should enhance the effect of the wrong 
scatterings. 

In Fig.~\ref{fig:vel} we compare the bare velocities and the dressed 
ones (dotted and continuous line, respectively). We show results only 
for $\alpha=1,2$. The effect of the dressing is clearly visible in the 
figure. Importantly, one consequence of the dressing is that the 
maximum of the velocities are shifted, as compared with the bare 
ones. This already implies that for some values of $\alpha,\lambda$ 
and $\beta,\mu$ one has that 
$\mathrm{sign}(v_{\alpha,\lambda}-v_{\beta,\mu})
\ne \mathrm{sign}(v_{\alpha,\lambda}^{\scriptscriptstyle\mathrm{bare}}
-v_{\beta,\mu}^{\scriptscriptstyle\mathrm{bare}})$, meaning that 
a priori the flea gas dynamics is not fully equivalent to the 
GHD. On the other hand, the behavior of the bare and dressed velocities 
is similar as a function of $\lambda$, suggesting that the effect of the 
wrong scatterings should be ``small''.

In Fig.~\ref{fig:coll} we investigate the effect of the wrong collisions 
on the shift of the quasiparticles trajectories. In panel (a) we show 
the average total shift $\Delta x$ experienced by the quasiparticle with 
rapidity $\lambda$ due to wrong scatterings (continuous line) and the 
total number of scatterings (dotted line). The results are for the 
quasiparticlew with $\alpha=1$, i.e., the unbound quasiparticles. 
We observe that the wrong scatterings have a small effect on $\Delta x$, 
which is barely visible in the figure. Similar behavior is observed 
for the two-particle bound states. The results are shown in 
panel (b). In panel (c) we show the effect of the wrong 
scatterings on the trajectories of the quasiparticles with $\alpha=1$. 
The figure shows the same data as in (a). The average shift due to the 
wrong scatterings is $\approx 10^{-3}$. 

Finally, two observations are in order. First, although panel 
(c) suggests a finite contribution of the wrong scattering to 
$\Delta x$, larger numerical simulations would be needed to 
ensure that the data are in the scaling limit $N,L\to\infty$ 
with $N$ the number of quasiparticles. Second, in principle, it 
should be possible to correct the effect of the wrong scatterings 
by imposing that upon colliding the displacement of the quasiparticle 
trajectories is the correct one $\mathrm{sign}(v_{\alpha,\lambda}-v_{\beta,\mu})$. 
Note that this requires knowing the dressed velocities $v_{\alpha,\lambda}$, 
which could be calculated during the simulation.

\subsection{Entanglement dynamics after a quench from inhomogeneous initial conditions}
\label{sec-ent-in}

Having established the validity of the flea gas method to simulate 
the entanglement dynamics after homogeneous quenches, we 
now consider the case of the inhomogeneous initial state in Fig.~\ref{fig0} 
(b). 
The calculation of the entanglement dynamics within the 
GHD framework is in general a complicated task. Explicit analytic results can be 
obtained only in few cases. For instance, the steady-state 
value of the von Neumann entanglement entropy for a finite subsystem placed next to 
the interface between the two chains (see Fig.~\ref{fig0} (b)) 
can be easily calculated. This corresponds to the limit $\ell/t\to 
0$. In this limit, the entire subsystem is described by the GGE with 
$\zeta=0$ (see section~\ref{ghd}). Following 
Ref.~\cite{AlCa17}, the density of the steady-state von Neumann entanglement entropy 
coincides with that of the  GGE entropy with $\zeta=0$. One has~\cite{Alba17} 
\begin{equation}
\label{steady-in}
S=\ell\sum_\alpha\int d\lambda  s_{\alpha,\lambda}(0). 
\end{equation}
Here $s_{\alpha,\lambda}(0)$ is the Yang-Yang entropy (cf.~\eqref{eq:s}) 
of the GGE with $\zeta=0$, which is obtained by using the GHD (see section~\ref{ghd}). 
The result does not depend on which side of the system one places the interval, 
as long as $\ell$ is finite. 
Interestingly, one can show that the $\zeta=0$  macrostate describes 
the entanglement growth at short times, i.e., the limit $1\ll t\ll\ell$ 
as well~\cite{Alba17,alba-2019}. 
First, the entanglement entropy is expected to grow linearly at short times. 
Here we refer to the slope of the linear growth as the entanglement 
production rate~\cite{Alba17,alba-2018,alba-2019}. 
The entanglement growth is due to the quasiparticles that cross the interface 
between the two chains. This suggests that the entanglement production rate is 
described by the $\zeta=0$ GGE. Indeed, if subsystem $A$ is the semi-infinite 
chain, the entanglement production rate $S/t$ is given as~\cite{alba-2019} 
\begin{equation}
\label{slope-in}
\frac{S}{t}=\sum_\alpha\int d\lambda \mathrm{sign}(\lambda) v_{\alpha,\lambda}(0)
s_{\alpha,\lambda}(0). 
\end{equation}
Here $v_{\alpha,\lambda}(0)$ is the group velocities of the particle-hole 
excitations around the $\zeta=0$ GGE, which are obtained from~\eqref{eq:vDoyon}. 
For a finite subsystem, the slope of the linear 
growth depends on the details of the bipartition. For simplicity we 
now consider the case of  
a finite interval of length $\ell$  placed in one of the two chains 
next to the interface (see Fig.~\ref{fig0} (b)). 
Clearly, the entanglement entropy gets contributions from 
both the edges of the subsystem. 
For short enough times but still in the linear regime, i.e, 
for large $t$ with $t/\ell\ll 1$, the 
contributions of the two edges decouple and can be summed independently. 
As in~\eqref{slope-in}, one of the edges of $A$ is described by the GGE with $\zeta=0$. 
On the other hand, the other one is described by the GGE with 
$\zeta=\pm\infty$, depending on which side subsystem 
$A$ is placed in. The entanglement production rate is given as 
\begin{equation}
\label{slope-in-1}
S=t\sum_\alpha\int d\lambda\Big[\mathrm{sign}(\lambda) v_{\alpha,\lambda}(0)
  s_{\alpha,\lambda}(0)+|v_{\alpha,\lambda}(\sigma\infty)
|s_{\alpha,\lambda}(\sigma\infty)\Big], 
\end{equation}
where $\sigma=\pm$ identifies the side in which subsystem $A$ is placed. 

\begin{figure}[t]
\begin{center}
\includegraphics*[width=0.65\linewidth]{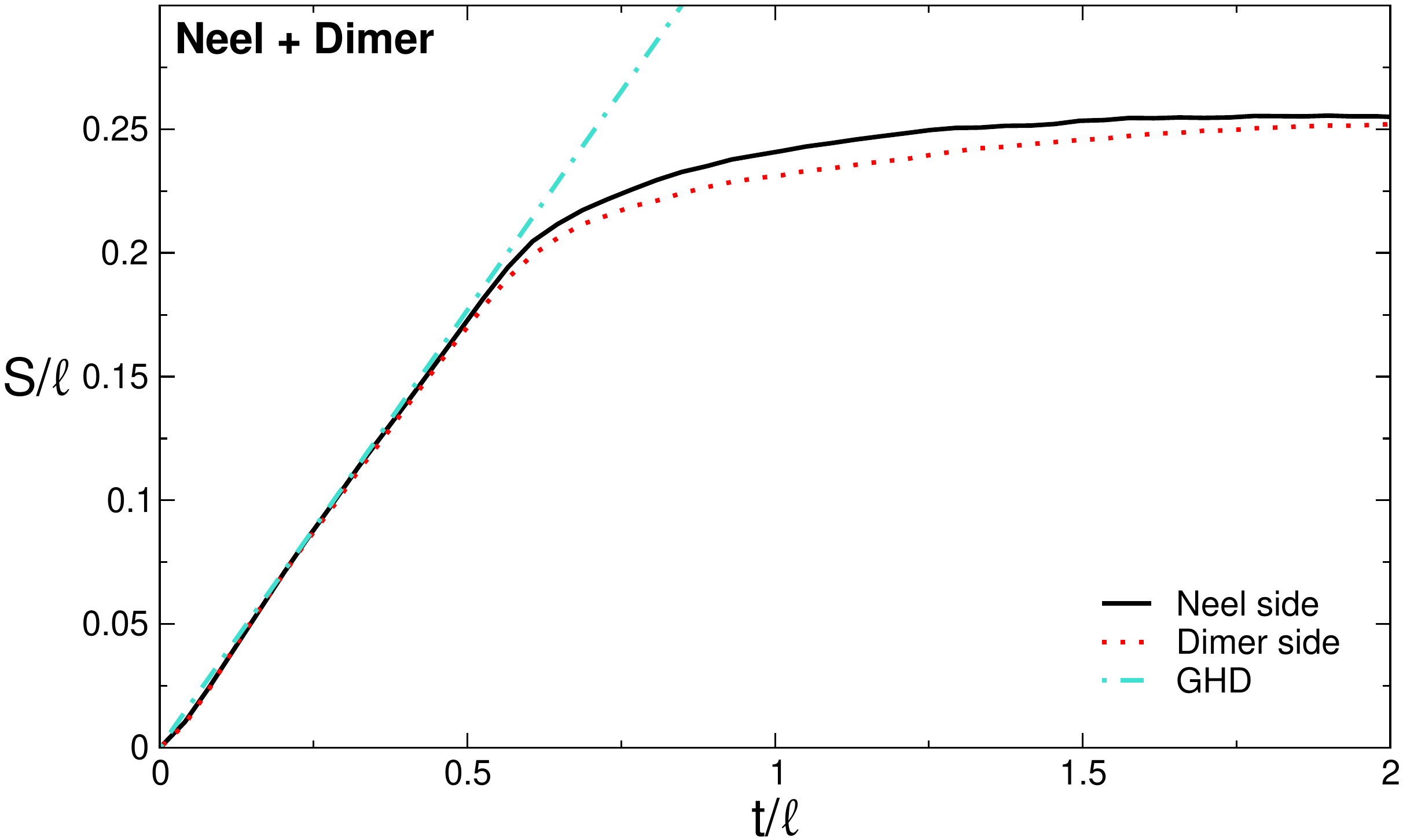}
\end{center}
\caption{ Dynamics of the von Neumann entanglement entropy after the 
 quench from the initial state obtained by joining the 
 N\'eel and the dimer state in the XXZ chain with 
 $\Delta=2$. The entropy density $S/\ell$ is plotted 
 versus the rescaled time $t/\ell$, with $\ell$ being the 
 subsystem size. The interval is placed next to the interface 
 between the two chains. Here we choose $\ell=100$, while the 
 total chain size is $L=2000$. The full and the dotted 
 lines are the flea gas results for an interval placed 
 in the N\'eel and dimer part, respectively. The results are 
 obtained by averaging over $\sim 10000$ realizations of the 
 dynamics. The dashed-dotted 
 line is the GHD prediction~\eqref{slope-in-1} valid in the space-time 
 scaling limit. Notice that the asymptotic value of the 
 entropy at $t\to\infty$ does not depend on the region where 
 the subsystem is placed. 
}
\label{fig2}
\end{figure}

To illustrate how these features emerge in the flea gas simulations, 
in Fig.~\ref{fig2} we present numerical results for the quench  
from the initial state obtained by joining the N\'eel state and the 
dimer state $|\mathrm{N}\rangle\otimes|\mathrm{D}\rangle$ 
(see section~\ref{sec-quenches} for the definition of these states, 
and section~\ref{sec-tba-quench} for their TBA treatment). The 
results are for the 
XXZ chain with $\Delta=2$. The full and dotted lines correspond to 
the bipartitions with interval $A$ being $[-\ell,0]$ (in the N\'eel region) 
and $[0,\ell]$ (in the dimer region) respectively. 
In both cases we consider $L=2000$ and $\ell=100$. 
The results are obtained by averaging over 
$10000$ realizations of the ``flea'' gas dynamics (see section~\ref{sec-flea}), and 
using~\eqref{eq:entropyFleaGas}. 

For the quench from $|\mathrm{N}\otimes \mathrm{dimer}
\rangle$, we observe that at $\Delta=2$ one has $s_{\alpha,\lambda}(+\infty)
\approx s_{\alpha,\lambda}(-\infty)$ and $v_{\alpha,\lambda}(+\infty)
\approx v_{\alpha,\lambda}(-\infty)$. From~\eqref{slope-in-1} one obtains that 
the entanglement production rate depends very mildly on which region the 
subsystem is placed. The theory predictions~\eqref{slope-in-1} for the entanglement 
production rates are not distinguishable on the scale of the figure and are 
reported as dashed-dotted line. 
%
%
At intermediate $\zeta=t/\ell$, the entanglement entropy depends on all the 
values of $\zeta$. This happens because the entangling quasiparticles explore 
macrostates with different $\zeta$ as they travel in subsystem $A$ (see Fig.~\ref{fig0}). 
Although it is possible, in principle, 
to write an analytic formula~\cite{alba-2019} for the evolution of 
the entanglement entropy at any time, its numerical evaluation is a 
demanding task. 
In contrast, the flea gas method allows to access easily 
the full-time entanglement dynamics, as it is clear from Fig.~\ref{fig2}. 

\begin{figure}[t]
\begin{center}
\includegraphics*[width=1\linewidth]{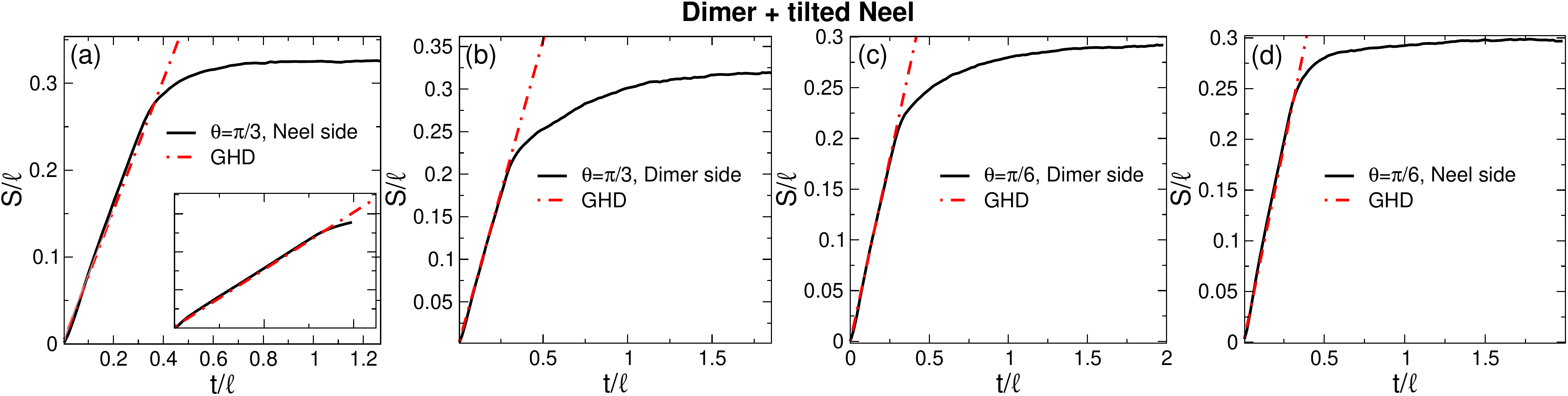}
\end{center}
\caption{ The same as in Fig.~\ref{fig2} for the quench 
 from the state obtained by joining the tilted N\'eel and 
 the dimer state. The data are for the XXZ chain at $\Delta=2$ 
 and for tilting angles $\theta=\pi/3$ (in (a) and (b)) and 
 $\theta=\pi/6$ (in (c) and (d)). The curves show the flea 
 gas results for a subsystem of size $\ell=100$ and chain 
 size $L=2000$. The data 
 are averaged over $\sim 10000$ realizations of the flea 
 gas dynamics. In (a) and (d) the subsystem is placed on the 
 N\'eel side, whereas in (b) and (c) it is in the dimer 
 side. The dashed-dotted lines are the GHD predictions 
 in the space-time scaling limit. The inset in (a) shows 
 results for $\ell=500$ and chain size $L=2000$. 
}
\label{fig3}
\end{figure}

In Fig.~\ref{fig3} we present further checks of the validity of 
the flea gas method for inhomogeneous quenches. We consider the 
initial state obtained by joining the 
tilted N\'eel state and the dimer state, i.e.,  $|\mathrm{N},\theta\rangle\otimes|\mathrm{dimer}\rangle$ 
(see section~\ref{sec-quenches}), where $\theta$ is the tilting angle. 
Panel (a) and (b) show results for $\theta=\pi/3$, whereas in (c) and (d) we consider 
$\theta=\pi/6$. In all the cases we choose $\ell=100$ and 
total system size  $L=2000$. In (a) and (d) the subsystem 
is placed on the N\'eel side ($A=[-\ell,0]$), whereas in (b) and (c)  is in the dimer 
side ($A=[0,\ell]$). The fact that the production rate depends on the 
position of the interval is now apparent. The dashed-dotted lines are the 
theory predictions (cf.~\eqref{slope-in-1}) 
for the entanglement production rates. 
In Fig.~\ref{fig3} (a) some deviations from~\eqref{slope-in-1} 
are visible. These, however,  are due to finite-size and 
finite-time effects. In the inset of Fig.~\ref{fig3} we report results for 
$\ell=500$, which are now in perfect agreement 
with~\eqref{slope-in-1}. We observe that in general very large 
subsystems are needed to provide a robust numerical 
check of the GHD prediction~\eqref{slope-in-1}. 
%
%

\subsection{Mutual information after quenches from inhomogeneous initial conditions}
\label{sec-ent-mi}

%
\begin{figure}[t]
\begin{center}
\includegraphics*[width=0.95\linewidth]{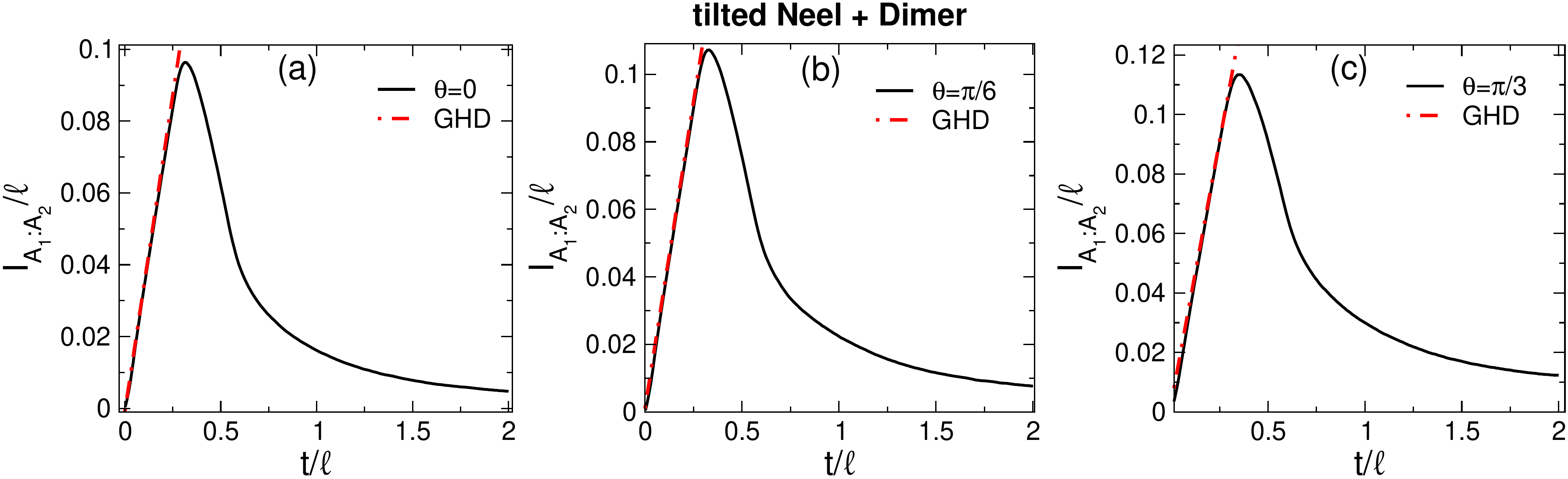}
\end{center}
\caption{ Dynamics of the mutual information between 
 two intervals after the quench from the state 
 $|\textrm{N\'eel},\theta\rangle\otimes|\textrm{dimer}\rangle$ 
 in the XXZ chain with $\Delta=2$. Here we show the 
 mutual information between two adjacent intervals next 
 to the interface between the two chains. The data are for 
 two intervals with $\ell=100$ embedded in a chain 
 with . The different panels 
 are for different tilting angles $\theta$. The dashed-dotted 
 lines are the GHD predictions for the slope of the linear 
 growth at short times. Notice that the slope depends on 
 the macrostate with $\zeta=0$ that describes the interface 
 between the two chains. 
}
\label{fig4}
\end{figure}
%

It is interesting to investigate the dynamics of the mutual information 
between two intervals. To this purpose, we now consider a tripartite system. Subsystem $A$ 
is made of two intervals $A_1$ and 
$A_2$ at a distance $d$. Here we consider only $d=0$, although the method works also 
for $d>0$. The two subsystems are embedded in an infinite system. 
The mutual information $I_{A_1:A_2}$ is a measure of the correlation 
shared between $A_1$ and $A_2$, although it  is not a proper 
measure of the entanglement between them. $I_{A_1:A_2}$ is defined as 
\begin{equation}
\label{mi-def}
I_{A_1:A_2}\equiv S_{A_1}+S_{A_2}-S_{A_1\cup A_2},
\end{equation}
where $S_{A_1}$, $S_{A_2}$, and $S_{A_1\cup A_2}$ are the von Neumann entanglement entropies of 
$A_1$, $A_2$ and $A_1\cup A_2$ with the rest of the system. 

In the quasiparticle picture, the mutual 
information is proportional to the entangled pairs that are shared only 
between $A_1$ and $A_2$. On the other hand, the contribution 
of the quasiparticles to the mutual information 
is, again, the GGE thermodynamic entropy. 
Thus, the flea gas formula for $I_{A_1:A_2}$ is the same 
as~\eqref{eq:entropyFleaGas} where the sum is restricted to 
the pairs of quasiparticles shared between $A_1$ and $A_2$. 

The qualitative behavior  of the mutual 
information is as follows. For two disjoint 
intervals at a distance $d$, the mutual information is zero at short times. 
At $t\sim d/t$, $I_{A_1:A_2}$ exhibits a linear increase. This corresponds 
to entangled pairs starting to be shared between $A_1$ and $A_2$. 
The growth persists up to $t\sim(d+\ell)/t$, when the 
mutual information starts to decrease. In systems with only one quasiparticle with 
perfect linear dispersion (as in CFT systems), 
the decrease is linear. In generic integrable models 
a much slower decrease is observed~\cite{alba-2018}. This is due to the fact that  
quasiparticles have a nontrivial dispersion, and  
slow quasiparticles entangle the two subsystems at long times. 
The mutual information can, in principle, be used as a tool to reveal the 
quasiparticle content of an integrable model. 
Typically, different species have different maximum velocities $v_{\alpha,M}$. 
This implies that if the distance between the two intervals is large enough, 
the mutual information exhibits a multi-peak 
structure in time, each peak corresponding to a different  
species~\cite{MBPC17,alba-2019a}. 

The mutual information after quenches from 
inhomogeneous initial states has not been investigated yet. In contrast with 
homogeneous global quenches~\cite{alba-2018}, 
deriving the quasiparticle picture for the mutual information in inhomogeneous settings 
is a formidable task. Again, the reason is that the quasiparticles trajectories are 
non trivial functions of time. 
We now show that the flea gas approach allows to simulate effectively the full-time 
dynamics of the mutual information. We 
restrict ourselves to the case of two adjacent intervals, although 
the method works for disjoint intervals as well. 

We present our results in Fig.~\ref{fig4}, focusing on the XXZ chain 
with $\Delta=2$. The initial state that we consider is $|\mathrm{N},
\theta\rangle\otimes|\mathrm{D}\rangle$. Different panels 
show different values of $\theta$. The data are for two equal-length 
adjacent intervals $[-\ell,0]$ and $[0,\ell]$  with $\ell=100$. The total chain length is $L=2000$. 
As expected, the mutual information 
is initially zero, it grows linearly at intermediate times, and it 
eventually decays to zero at asymptotically long times. Importantly, 
the initial slope of the mutual information depends only on the GGE with $\zeta=0$ 
because the interface between $A_1$ and $A_2$ is at  the origin. 
In particular, the slope of the inital growth of the mutual information 
coincides with the entanglement production rate 
for the two semi-infinite chains (see Eq.~\eqref{slope-in}). 
This initial growth is reported in Fig.~\ref{fig4} as dashed-dotted 
line, and it perfectly describes the behavior of the flea gas results. 

\section{Conclusions}
\label{sec-con}

In this work we showed that the so-called flea gas method put forward in Ref.~\cite{DDKY17a} 
provides a versatile tool  for 
simulating the entanglement dynamics after quenches from generic initial 
states in integrable systems. We benchmarked the method in the Heisenberg 
XXZ chain, although it can be applied, in principle, to any integrable model. 
The method works for arbitrary initial states, both globally homogeneous 
as well as piecewise homogeneous. For globally homogeneous quenches the approach 
requires only the GGE macrostate that describes the steady-state. 
For piecewise homogeneous states, the key ingredients are the GGE macrostates describing 
the steady-state in the bulk of the two systems. Although in this case the 
entanglement dynamics can be obtained, in principle, by combining 
the quasiparticle picture with the Generalized Hydrodynamics, 
obtaining explicit formulas~\cite{alba-2019} is a 
demanding task because the trajectories of the quasiparticles are non trivial functions 
of time. Indeed, results can be obtained only in some limits. In contrast, 
in this work we showed that the flea gas approach allows to obtain easily the full-time dynamics of the 
entanglement entropy and of the mutual information between two intervals. 
Thus, the method paves the way to the study of entanglement dynamics 
using ``molecular dynamics'' simulations. 

Our results open several possible new research directions. 
First, it would be important to investigate whether it is possible to prove 
analytically that for the XXZ chain the flea gas dynamics as described in 
section~\ref{sec-flea-xxz} gives the correct dressing for the group velocities. 

Also, it would be useful to apply the method to more complicated 
setups, such as multipartite systems, or different initial states. 
Also, it would be important to go beyond the ballistic regime, studying corrections to the linear 
entanglement growth. This requires first to understand the subleading diffusive 
corrections in the flea gas method. Second, it requires to modify 
the flea gas dynamics to correctly reproduce the diffusive corrections that arise from 
the quantum fluctuations~\cite{DBD,de-nardis-2018a}. 
An interesting direction would be to generalize the flea gas approach 
to study  the entanglement dynamics in the presence of defects or impurities. 
Finally, it would be 
enlightening to understand whether it is possible to treat the entanglement 
of operators in integrable spin chains by using the flea gas approach, generalizing 
the results of Ref.~\cite{bobenko} for the Rule $54$ chain. 

\section{Acknowledgements}
MM acknowledges support from ERC under Consolidator grant number 771536 (NEMO). 
VA acknowledges support from the D-ITP consortium, a program of the NWO. 
VA also acknowledges support  from  the  European  Research  Council  under  ERC 
Advanced grant 743032 DYNAMINT. 


\nolinenumbers

\begin{thebibliography}{99}
\bibitem{ge-15} 
  C.~Gogolin and J.~Eisert, {\it Equilibration, thermalisation, and the emergence of statistical mechanics in closed quantum systems,} Rep. Prog. Phys. {\bf 79}, 056001 (2016), \doi{10.1088/0034-4885/79/5/056001}. 

\bibitem{R08}  
  M. Rigol, V. Dunjko, and M. Olshanii, 
  {\it Thermalization and its mechanism for generic isolated quantum systems}, { Nature {\bf 452}, 854 (2008),}
\doi{10.1038/nature06838}.

\bibitem{ef-16}
  F. H. L. Essler and M. Fagotti, {\it Quench dynamics and relaxation in isolated integrable quantum spin chains},
  J. Stat. Mech. P064002 (2016),
\doi{10.1088/1742-5468/2016/06/064002}.

\bibitem{ViRi16} 
L. Vidmar and M. Rigol, 
{\it Generalized Gibbs ensemble in integrable lattice models,}
J. Stat. Mech. P064007 (2016) ,
\doi{10.1088/1742-5468/2016/06/064007}.

\bibitem{CQA16}
J-S. Caux, 
{\it The Quench Action,}
J. Stat. Mech. P064006 (2016),
\doi{10.1088/1742-5468/2016/06/064006}.



\bibitem{DKPR16} 
L. D'Alessio, Y. Kafri, A. Polkovnikov, and M. Rigol, 
{\it From quantum chaos and eigenstate thermalization to statistical mechanics and thermodynamics,}
Adv. Phys. {\bf 65}, 239 (2016),
\doi{10.1080/00018732.2016.1198134}.


\bibitem{FaCa08} 
M. Fagotti and P. Calabrese, 
{\it Evolution of entanglement entropy following a quantum quench: Analytic results for the XY chain in a transverse magnetic field,}
Phys. Rev. A {\bf 78}, 10306 (2008),
\doi{10.1103/PhysRevA.78.010306}.

\bibitem{CCsemiclassics} 
P.~Calabrese and J.~Cardy,
{\it Evolution of entanglement entropy in one-dimensional systems,}
J. Stat. Mech. P04010 (2005),
\doi{10.1088/1742-5468/2005/04/P04010}.

\bibitem{dmcf-06}
G. De Chiara, S. Montangero, P. Calabrese, and R. Fazio, 
{\it Entanglement entropy dynamics of Heisenberg chains,}
J. Stat. Mech. (2006) P03001,
\doi{10.1088/1742-5468/2006/03/P03001}.

\bibitem{lauchli-2008}
  A. L{\" a}uchli and C. Kollath,
{\it Spreading of correlations and entanglement after a quench in the one-dimensional Bose–Hubbard model,}
J. Stat. Mech.  P05018 (2008),
\doi{10.1088/1742-5468/2008/05/P05018}.

\bibitem{ep-08}
V. Eisler and I. Peschel, 
{\it Entanglement in a periodic quench},
Ann. Phys. {\bf 17}, 410 (2008),
\doi{10.1002/andp.200810299}.

\bibitem{hk-13}
H.~Kim and D.~A.~Huse, 
{\it Ballistic spreading of entanglement in a diffusive nonintegrable system},
Phys.\  Rev.\ Lett.\ {\bf111}, 127205 (2013),
\doi{10.1103/PhysRevLett.111.127205}.

\bibitem{Gura13} 
V. Gurarie,
{\it Global large time dynamics and the generalized Gibbs ensemble},
J. Stat. Mech. P02014 (2013),
\doi{10.1088/1742-5468/2013/02/P02014}.

\bibitem{Fago13}
M. Fagotti, 
{\it Finite-size corrections versus relaxation after a sudden quench},
Phys. Rev. B {\bf 87}, 165106 (2013),
\doi{10.1103/PhysRevB.87.165106}.


\bibitem{ctc-14} 
A. Coser, E. Tonni, and P. Calabrese,
{\it Entanglement negativity after a global quantum quench},
J. Stat. Mech.  (2014) P12017
\doi{10.1088/1742-5468/2014/12/P12017}.

\bibitem{nr-14}
M. G.~Nezhadhaghighi and M. A. Rajabpour, 
{\it Entanglement dynamics in short- and long-range harmonic oscillators},
Phys. Rev. B {\bf 90}, 205438 (2014),
\doi{10.1103/PhysRevB.90.205438}.

\bibitem{buyskikh-2016}
A. S. Buyskikh, M. Fagotti, J. Schachenmayer, F. Essler, and A. J. Daley,
{\it Entanglement growth and correlation spreading with variable-range interactions in spin and fermionic tunneling models},
Phys. Rev. A {\bf 93}, 053620,
\doi{https://link.aps.org/doi/10.1103/PhysRevA.93.053620}.


\bibitem{cotler-2016}
J.~S.~Cotler, M.~P.~Hertzberg, M.~Mezei, and M.~T.~Mueller,
{\it Entanglement growth after a global quench in free scalar field theory},
JHEP {\bf 11}, 166 (2016),
\doi{10.1007/JHEP11(2016)166}. 

\bibitem{kctc-17}
M. Kormos, M. Collura, G. Tak\'acs, and P. Calabrese,
{\it Real-time confinement following a quantum quench to a non-integrable model},
Nature Phys. {\bf 13}, 246 (2017),
\doi{10.1038/nphys3934}.

\bibitem{MBPC17} 
M. Mesty\'an, B. Bertini, L. Piroli, and P. Calabrese, 
{\it Exact solution for the quench dynamics of a nested integrable system},
J. Stat. Mech. P83103 (2017),
\doi{10.1088/1742-5468/aa7df0}.

\bibitem{mkz-17}
C. Pascu Moca, M. Kormos, and G. Zarand, 
{\it Hybrid Semiclassical Theory of Quantum Quenches in One-Dimensional Systems},
Phys. Rev. Lett. {\bf 119}, 100603 (2017),
\doi{10.1103/PhysRevLett.119.100603}.

\bibitem{p-18}
P. Calabrese,
  {\it Entanglement and thermodynamics in non-equilibrium isolated quantum systems},
J. Phys. A  {\bf 504},  31-44 (2018),
 \doi{10.1016/j.physa.2017.10.011}.

\bibitem{fnr-17}
I. Frerot, P. Naldesi, and T. Roscilde,  
{\it Multispeed prethermalization in quantum spin models with power-law decaying interactions},
Phys. Rev. Lett. {\bf 120}, 050401 (2018),
\doi{10.1103/PhysRevLett.120.050401}.

\bibitem{ckt-18}
M. Collura, M. Kormos, and G. Takacs, 
{\it Dynamical manifestation of the Gibbs paradox after a quantum quench},
Phys. Rev. A {\bf 98}, 053610 (2018),
\doi{10.1103/PhysRevA.98.053610}.

\bibitem{d-17}
J. Dubail, 
{\it Entanglement scaling of operators: a conformal field theory approach, with a glimpse of simulability of long-time dynamics in 1+1d},
J. Phys. A {\bf 50}, 234001 (2017),
\doi{10.1088/1751-8121/aa6f38}.

\bibitem{BTC:beyondpairs}
B.~Bertini, E.~Tartaglia, and P.~Calabrese, 
{\it Entanglement and diagonal entropies after a quench with no pair structure},
J. Stat. Mech.  063104 (2018),
\doi{10.1088/1742-5468/aac73f}.

\bibitem{BC:beyondpairs2}
A.~Bastianello, P.~Calabrese,
{\it Spreading of entanglement and correlations after a quench with intertwined quasiparticles}, 
SciPost Phys. {\bf 5}, 033 (2018),
\doi{10.21468/SciPostPhys.5.4.033}.

\bibitem{LS:CFT} 
H. Liu and S. J. Suh, 
{\it Entanglement Tsunami: Universal Scaling in Holographic Thermalization},
Phys. Rev. Lett. {\bf 112}, 011601 (2014),
\doi{10.1103/PhysRevLett.112.011601}.

\bibitem{CLM:minimalcut} 
H. Casini, H. Liu, and M. Mezei, 
{\it Spread of entanglement and causality},
J. High Energy Phys. {\bf 07} (2016) 077,
\doi{10.1007/JHEP07(2016)077}.

\bibitem{ABGH:CFT}
C.T. Asplund, A. Bernamonti, F. Galli, T. Hartman, 
{\it Entanglement scrambling in 2d conformal field theory},
J. High Energ. Phys. {\bf 09} (2015) 11,
\doi{10.1007/JHEP09(2015)110}. 

\bibitem{LM:CFT}
S. Leichenauer and M. Moosa, 
  {\it Entanglement tsunami in (1+1)-dimensions}
Phys. Rev. D {\bf 92}, 126004 (2015),
 \doi{10.1103/PhysRevD.92.126004}.

\bibitem{AlCa17}
V. Alba and P. Calabrese,
{\it Entanglement and thermodynamics after a quantum quench in integrable systems},
PNAS {\bf 114}, 7947 (2017),
\doi{10.1073/pnas.1703516114}.

\bibitem{nahum-17}
A. Nahum, J. Ruhman, S. Vijay, and J. Haah,
{\it Quantum entanglement growth under random unitary dynamics},
Phy. Rev. X {\bf 7}, 031016 (2017),
\doi{10.1103/PhysRevX.7.031016}.

\bibitem{BKP:entropy}
B. Bertini, P. Kos, and T. Prosen, 
{\it Entanglement Spreading in a Minimal Model of Maximal Many-Body Quantum Chaos},
Phys. Rev. X {\bf 9}, 021033 (2019),
\doi{10.1103/PhysRevX.9.021033}

\bibitem{MM}
M. R. M. Mozaffar and A. J. Mollabashi, 
{\it Entanglement evolution in Lifshitz-type scalar theories},
J. High Energ. Phys. {\bf 2019} (2019) 137,
\doi{10.1007/JHEP01(2019)137}. 

\bibitem{area}
J.~Eisert, M.~Cramer, and M.~B.~Plenio,
{\it Colloquium: Area laws for the entanglement entropy},
Rev.\ Mod.\ Phys.\ {\bf 82}, 277 (2010),
\doi{10.1103/RevModPhys.82.277}. 

\bibitem{amico-2008}
L.~Amico, R.~Fazio, A.~Osterloh, V.~Vedral, 
{\it Entanglement in many-body systems},
Rev.\ Mod.\ Phys.\ {\bf 80}, 517 (2008),
\doi{10.1103/RevModPhys.80.517}.

\bibitem{calabrese-2009} 
P.~Calabrese, J.~Cardy, and B.~Doyon, 
{\it Introduction to 'Entanglement entropy in extended quantum systems'},
J. Phys. A {\bf 42} 500301 (2009),
\doi{10.1088/1751-8121/42/50/500301}.

\bibitem{laflorencie-2016}
N.~Laflorencie, 
{\it Quantum entanglement in condensed matter systems},
Phys. Rep. {\bf 646}, 1 (2016),
\doi{10.1016/j.physrep.2016.06.008}. 


V. Alba and P. Calabrese, 
{\it Entanglement dynamics after quantum quenches in generic integrable  systems},
SciPost Phys. {\bf 4}, 017 (2018),
\doi{10.21468/SciPostPhys.4.3.017}.

\bibitem{rigol-2007}
M. Rigol, V. Dunjko, V. Yurovsky, and M. Olshanii, 
{\it Relaxation in a completely integrable many-body quantum system: an ab initiostudy of the dynamics of the highly excited states of 1d lattice hard-core bosons},
Phys. Rev. Lett. {\bf 98}, 050405 (2007),
\doi{10.1103/PhysRevLett.98.050405}.


\bibitem{wouters-2014}
B.~Wouters, J. De Nardis, M. Brockmann, D. Fioretto, M. Rigol, and J.-S. Caux, 
{\it Quenching the anisotropic Heisenberg chain: exact solution and generalized Gibbs ensemble predictions},
Phys. Rev. Lett. {\bf 113}, 117202 (2014),
\doi{10.1103/PhysRevLett.113.117202}.

\bibitem{pozsgay-2014}
B.~Pozsgay, M.~Mesty\'an, M.~A.~Werner, M.~Kormos, G.~Zar\'and, and G.~Tak\'acs, 
{\it Correlations after quantum quenches in the XXZ spin chain: failure of the generalized Gibbs ensemble},
Phys. Rev. Lett. {\bf 113}, 117203 (2014),
\doi{10.1103/PhysRevLett.113.117203}. 


\bibitem{IDWC15} 
E. Ilievski, J. De Nardis, B. Wouters, J.-S. Caux, F. H. L. Essler, and T. Prosen, 
{\it Complete generalized Gibbs ensembles in an interacting theory},
Phys. Rev. Lett. {\bf 115}, 157201 (2015),
\doi{10.1103/PhysRevLett.115.157201}.



\bibitem{BEL:lightcone}
L. Bonnes, F. H. L. Essler, A. M. L\"auchli, 
{\it “Light-cone” dynamics after quantum quenches in spin chains},
Phys. Rev. Lett. {\bf 113}, 187203 (2014),
\doi{10.1103/PhysRevLett.113.187203}.


\bibitem{ac-17b}
V.~Alba and P.~Calabrese, 
{\it Quench action and Rényi entropies in integrable systems},
Phys. Rev. B {\bf 96}, 115421 (2017),
\doi{10.1103/PhysRevB.96.115421};\\
V.~Alba and P.~Calabrese, 
{\it Rényi entropies after releasing the Néel state in the XXZ spin-chain},
J.\ Stat.\ Mech.\ (2017) 113105,
\doi{10.1088/1742-5468/aa934c}.


\bibitem{alba-2018}
V.~Alba,
{\it Towards a generalized hydrodynamics description of Rényi entropies in integrable systems},
Phys. Rev. B {\bf 99}, 045150 (2019),
\doi{10.1103/PhysRevB.99.045150}. 


\bibitem{mestyan-2018}
M.~Mesty\'an, V.~Alba, and P.~Calabrese, 
{\it Rényi entropies of generic thermodynamic macrostates in integrable systems},
J.\ Stat.\ Mech.\ (2018) 083104,
\doi{10.1088/1742-5468/aad6b9}.



\bibitem{alba-2018a}
V.~Alba and P.~Calabrese, 
{\it Quantum information dynamics in multipartite integrable systems},
\href{https://arxiv.org/abs/1809.09119}{arXiv:1809.09119 (2018)}. 


\bibitem{BF16}
B. Bertini and M. Fagotti, 
{\it Determination of the nonequilibrium steady state emerging from a defect},
Phys. Rev. Lett. {\bf 117}, 130402 (2016),
\doi{10.1103/PhysRevLett.117.130402}.


\bibitem{BCDF16} 
B.~Bertini, M.~Collura, J.~De~Nardis, and M.~Fagotti, 
{\it Transport in out-of-oquilibrium XXZ chains: exact profiles of charges and currents},
Phys. Rev. Lett. {\bf 117}, 207201 (2016),
\doi{10.1103/PhysRevLett.117.207201}.


\bibitem{CaDY16} 
O. A. Castro-Alvaredo, B. Doyon, and T. Yoshimura, 
{\it Emergent Hydrodynamics in Integrable Quantum Systems Out of Equilibrium},
Phys. Rev. X {\bf 6}, 41065 (2016),
\doi{10.1103/PhysRevX.6.041065}.

\bibitem{Alba17} 
V. Alba, 
{\it Entanglement and quantum transport in integrable systems},
Phys. Rev. B {\bf 97}, 245135 (2018),
\doi{10.1103/PhysRevB.97.245135}.

\bibitem{BFPC}
B. Bertini, M. Fagotti, L. Piroli, and P. Calabrese, 
{\it Entanglement evolution and generalised hydrodynamics: noninteracting systems},
J. Phys. A: Math. Theor. {\bf 51}, 39LT01 (2018),
\doi{10.1088/1751-8121/aad82e}.

\bibitem{alba-2019}
V.~Alba, B.~Bertini, and M.~Fagotti,
{\it Entanglement evolution and generalised hydrodynamics: interacting integrable systems},
\href{https://arxiv.org/abs/1903.00467}{arXiv:1903.00467} (2019). 


\bibitem{DDKY17} 
B.~Doyon, J.~Dubail, R.~Konik, and T.~Yoshimura, 
{\it Large-scale description of interacting one-dimensional Bose gases: generalized hydrodynamics supersedes conventional hydrodynamics},
Phys. Rev. Lett. {\bf 119}, 195301 (2017),
\doi{10.1103/PhysRevLett.119.195301}. 

\bibitem{jerome-2018}
M.~Schemmer, I.~Bouchoule, B.~Doyon, and J.~Dubail,
{\it Generalized Hydrodynamics on an Atom Chip},
Phys. Rev. Lett. {\bf 122}, 090601 (2019),
\doi{10.1103/PhysRevLett.122.090601}.


\bibitem{bdwy-18}
A. Bastianello, B. Doyon, G. Watts, and T. Yoshimura,
{\it Generalized hydrodynamics of classical integrable field theory: the sinh-Gordon model},
SciPost Phys. {\bf 4}, 045 (2018),
\doi{10.21468/SciPostPhys.4.6.045}.


\bibitem{DoYo17a}
B.~Doyon and H.~Spohn, 
{\it Dynamics of hard rods with initial domain wall state},
J. Stat. Mech.  073210 (2017),
\doi{10.1088/1742-5468/aa7abf}.

\bibitem{Taka99}
	M.~Takahashi, {\it Thermodynamics of one-dimensional solvable models}, Cambridge University Press, Cambridge, 1999.


\bibitem{kauf}
A. M. Kaufman, M. E. Tai, A. Lukin, M. Rispoli, R. Schittko, P. M. Preiss, and M. Greiner,
{\it Quantum thermalization through entanglement in an isolated many-body system},
Science {\bf 353}, 794  (2016),
\doi{10.1126/science.aaf6725}.


\bibitem{dls-13}
J. M. Deutsch, H. Li, and A. Sharma, 
{\it Microscopic origin of thermodynamic entropy in isolated systems},
Phys. Rev. E {\bf 87}, 042135 (2013),
\doi{10.1103/PhysRevE.87.042135}.

\bibitem{collura-2014}
M.~Collura, M.~Kormos, and P.~Calabrese, 
{\it Stationary entanglement entropies following an interaction quench in 1D Bose gas},
J. Stat. Mech. P01009 (2014),
\doi{10.1088/1742-5468/2014/01/P01009};

A. Nahum,  S. Vijay, and J. Haah, 
{\it Operator Spreading in Random Unitary Circuits},
Phys. Rev. X {\bf 8}, 021014 (2018),
\doi{10.1103/PhysRevX.8.021014};

C. Jonay, D. A. Huse, and A. Nahum,
{\it Coarse-grained dynamics of operator and state entanglement},
\href{https://arxiv.org/abs/1803.00089}{arXiv:1803.00089} (2019).

\bibitem{nwfs-18}
Y. O. Nakagawa, M. Watanabe, H. Fujita, and S. Sugiura, 
{\it Universality in volume-law entanglement of scrambled pure quantum states},
Nat. Comm. {\bf 9}, 1635 (2018),
\doi{10.1038/s41467-018-03883-9}.

\bibitem{ggemc}
A.~Cassidy, C.~W.~Clark, and M.~Rigol, 
{\it Generalized thermalization in an integrable lattice system},
Phys.\ Rev.\ Lett.\ {\bf 106}, 140405 (2011),
\doi{10.1103/PhysRevLett.106.140405}.


\bibitem{string-charge}
E.~Ilievski, E.~Quinn, J.~De~Nardis, and M.~Brockmann,
{\it String-charge duality in integrable lattice models},
J. Stat. Mech.  063101 (2016),
\doi{10.1088/1742-5468/2016/06/063101}. 

\bibitem{suzuki-99}
J.~Suzuki, 
{\it Spinons in magnetic chains of arbitrary spins at finite temperatures},
J. Phys. A {\bf 32}, 2341 (1999),
\doi{10.1088/0305-4470/32/12/008}.


\bibitem{piroli-2016}
L.~Piroli, E.~Vernier, and P.~Calabrese, 
{\it Exact steady states for quantum quenches in integrable Heisenberg spin chains},
Phys. Rev. B {\bf 94}, 054313 (2016),
\doi{10.1103/PhysRevB.94.054313}. 


\bibitem{DDKY17b} 
B.~Doyon and H.~Spohn, 
{\it Drude Weight for the Lieb-Liniger Bose Gas},
SciPost Phys. {\bf 3}, 039 (2017),
\doi{10.21468/SciPostPhys.3.6.039}.

\bibitem{BePi17} 
B. Bertini and L. Piroli, 
{\it Low-temperature transport in out-of-equilibrium XXZ chains},
J. Stat. Mech. 033104 (2018),
\doi{10.1088/1742-5468/aab04b}.

\bibitem{Bulchandani-17}
V. B. Bulchandani, 
{\it On classical integrability of the hydrodynamics of quantum integrable systems},
J. Phys. A {\bf 50}, 435203, 2017,
\doi{10.1088/1751-8121/aa8c62}.

\bibitem{DoYo17} 
B.~Doyon and T.~Yoshimura, 
{\it A note on generalized hydrodynamics: inhomogeneous fields and other  concepts},
SciPost Phys. {\bf 2}, 014 (2017),
\doi{10.21468/SciPostPhys.2.2.014}.

\bibitem{IlDe17} 
E. Ilievski and J. De Nardis, 
{\it Microscopic origin of ideal conductivity in integrable quantum models},
Phys. Rev. Lett. {\bf 119}, 020602 (2017),
\doi{10.1103/PhysRevLett.119.020602};

E. Ilievski and J. De Nardis, 
{\it Ballistic transport in the one-dimensional Hubbard model: the hydrodynamic approach},
Phys. Rev. B {\bf 96}, 081118(R) (2017),
\doi{10.1103/PhysRevB.96.081118}.

\bibitem{BVKM17} 
V. B. Bulchandani, R. Vasseur, C. Karrasch, and J. E. Moore, 
{\it Bethe-Boltzmann hydrodynamics and spin transport in the XXZ chain},
Phys. Rev. B {\bf 97}, 045407 (2018),
\doi{10.1103/PhysRevB.97.045407};\\
V. B. Bulchandani, R. Vasseur, C. Karrasch, and J. E. Moore, 
{\it Solvable hydrodynamics of quantum integrable systems},
Phys. Rev. Lett. {\bf 119}, 220604 (2017),
\doi{10.1103/PhysRevLett.119.220604}.

\bibitem{cdv-17}
A. De Luca, M. Collura, and J. De Nardis,
{\it Nonequilibrium spin transport in integrable spin chains: Persistent currents and emergence of magnetic domains},
Phys. Rev. B {\bf 96}, 020403(R)  (2017),
\doi{10.1103/PhysRevB.96.020403};\\
M. Collura, A. De Luca, and J. Viti, 
{\it Analytic solution of the domain-wall nonequilibrium stationary state},
Phys. Rev. B {\bf 97}, 081111 (2018),
\doi{10.1103/PhysRevB.97.081111}.

\bibitem{BePC18} 
B. Bertini, L. Piroli, and P. Calabrese, 
{\it Universal broadening of the light cone in low-temperature transport},
Phys. Rev. Lett. {\bf 120}, 176801 (2018),
\doi{10.1103/PhysRevLett.120.176801}.

\bibitem{PDCB17} 
L. Piroli, J. De Nardis, M. Collura, B. Bertini, and M. Fagotti, 
{\it Transport in out-of-equilibrium XXZ chains: nonballistic behavior and correlation functions},
Phys. Rev. B {\bf 96}, 115124 (2017),
\doi{10.1103/PhysRevB.96.115124}.

\bibitem{DSY17} 
B.~Doyon, H.~Spohn, and T.~Yoshimura, 
{\it A geometric viewpoint on generalized hydrodynamics},
Nucl. Phys. B {\bf 926}, 570 (2017),
\doi{10.1016/j.nuclphysb.2017.12.002}.

\bibitem{MBPC18}
M.~Mesty\'an, B. Bertini, L. Piroli, and P. Calabrese, 
{\it Spin-charge separation effects in the low-temperature transport of one-dimensional Fermi gases},
Phys. Rev. B {\bf 99}, 014305 (2019),
\doi{10.1103/PhysRevB.99.014305}. 


\bibitem{mvc-18}
L. Mazza, J. Viti, M. Carrega, D. Rossini, and A. De Luca, 
{\it Energy transport in an integrable parafermionic chain via generalized hydrodynamics},
Phys. Rev. B {\bf 98}, 075421,
\doi{10.1103/PhysRevB.98.075421}.

\bibitem{bertini-2019}
B.~Bertini, L.~Piroli, and M.~Kormos,  
{\it Transport in the sine-Gordon field theory: from generalized hydrodynamics to semiclassics},
\href{https://arxiv.org/abs/1904.02696}{arXiv:1904.02696} (2019). 


\bibitem{spohn-2019}
H.~Spohn, 
{\it Generalized Gibbs ensembles of the classical Toda chain},
\href{https://arxiv.org/abs/1902.07751}{arXiv:1902.07751} (2019). 

\bibitem{doyon-2019}
B.~Doyon and J.~Myers, 
{\it Fluctuations in ballistic transport from Euler hydrodynamics},
\href{https://arxiv.org/abs/1902.00320}{arXiv:1902.00320} (2019).


\bibitem{myers-2019}
J.~Myers, M.~J.~Bhaseen, R.~J.~Harris, and B.~Doyon, 
{\it Transport fluctuations in integrable models out of equilibrium},
\href{https://arxiv.org/abs/1812.02082}{arXiv:1812.02082} (2018).


\bibitem{alvise-2018}
A.~Bastianello and A.~De Luca, 
{\it Integrability-protected adiabatic reversibility in quantum spin chains},
\href{https://arxiv.org/abs/1811.07922}{arXiv:1811:07922} (2018).

\bibitem{vu-2018}
D.~L.~Vu and T.~Yoshimura, 
{\it Equations of state in generalized hydrodynamics},
SciPost Phys. {\bf 6}, 023 (2019),
\doi{10.21468/SciPostPhys.6.2.023}.


\bibitem{mazza-2018}
L.~Mazza, J.~Viti, M.~Carrega, D.~Rossini, and A.~De Luca, 
{\it Energy transport in an integrable parafermionic chain via generalized hydrodynamics},
Phys. Rev. B {\bf 98}, 075421 (2018),
\doi{10.1103/PhysRevB.98.075421}.





\bibitem{null}
M.~Fagotti, 
{\it Higher-order generalized hydrodynamics in one dimension: the noninteracting test},
Phys.\ Rev.\ B {\bf 96}, 220302(R)  (2017),
\doi{10.1103/PhysRevB.96.220302}.


\bibitem{DBD}
J.~De Nardis, D.~Bernard, B.~Doyon, 
{\it Hydrodynamic Diffusion in Integrable Systems},
Phys. Rev. Lett. {\bf 121}, 160603 (2018),
\doi{10.1103/PhysRevLett.121.160603}.

\bibitem{de-nardis-2018a}
J.~ De Nardis, D.~Bernard, and B.~Doyon,
{\it Diffusion in generalized hydrodynamics and quasiparticle scattering},
SciPost Phys. 6, 049 (2019),
\doi{10.21468/SciPostPhys.6.4.049}.

\bibitem{GHKV}
S.~Gopalakrishnan, D.~A. Huse, V.~Khemani, R.~Vasseur, 
{\it Hydrodynamics of operator spreading and quasiparticle diffusion in interacting integrable systems},
Phys. Rev. B {\bf 98}, 220303(R) (2019),
\doi{10.1103/PhysRevB.98.220303}.

\bibitem{sarang-2018a}
S.~Gopalakrishnan and R.~Vasseur, 
{\it Kinetic Theory of Spin Diffusion and Superdiffusion in XXZ Spin Chains},
Phys. Rev. Lett. {\bf 122}, 127202
\doi{10.1103/PhysRevLett.122.127202}.

\bibitem{DDKY17a} 
B.~Doyon, T.~Yoshimura, and J.-S.~Caux, 
{\it Soliton Gases and Generalized Hydrodynamics},
Phys. Rev. Lett. {\bf 120}, 045301 (2018),
\doi{10.1103/PhysRevLett.120.045301}.

\bibitem{boldrighini-1983}
C.~Boldrighini, R.~L.~Dobrushin,  and  Yu.~M.~Sukhov,  
{\it One-dimensional hard rod caricature of hydrodynamics},
J. Stat. Phys. {\bf 31}, 577 (1983),
\doi{10.1007/BF01019499}. 


\bibitem{smac}
W.~Krauth, \emph{Statistical Mechanics: Algorithms and Computation}, Oxford University Press, Oxford, 2006. 

\bibitem{alba-2019a}
V.~Alba and P.~Calabrese,
  {\it Quantum information scrambling after a quantum quench},
 \href{https://arxiv.org/abs/1903.09176}{arXiv:1903.09176} (2019). 

\bibitem{bobenko}
V.~Alba, J.~Dubail, and M.~Medenjak, 
{\it Operator entanglement in interacting integrable quantum systems: the case of the rule 54 chain},
\href{https://arxiv.org/abs/1901.04521}{arXiv:1901.04521} (2019). 


%





%


















%

%

%


%

%



%

%

\end{thebibliography}
\end{document}